\documentclass[twocolumn,preprintnumbers,amsmath,amssymb,aps,prb,longbibliography]{revtex4-1}
\usepackage{graphicx}
\usepackage{color}
\usepackage{dcolumn}
\usepackage{bm}

\begin{document}

\title{Directional Locking and Hysteresis in Stripe and Bubble Forming Systems on One-Dimensional Periodic Substrates with a Rotating Drive
}
\author{
C. Reichhardt and C. J. O. Reichhardt 
} 
\affiliation{
Theoretical Division and Center for Nonlinear Studies,
Los Alamos National Laboratory, Los Alamos, New Mexico 87545, USA
}
\date{\today}

\begin{abstract}
We examine the dynamics of a two-dimensional stripe, bubble, and crystal forming system interacting with a periodic one-dimensional substrate under an applied drive that is rotated with respect to the substrate periodicity direction $x$.
We find that the stripes remain strongly directionally locked to the $x$ direction for an extended range of drives before undergoing motion parallel to the drive. 
In some cases, the stripes break apart at the unlocking transition, but
can dynamically reform into stripes aligned
perpendicular to the $x$ direction,
producing hysteresis
in the directional locking and unlocking transitions.
In contrast, moving anisotropic crystal and bubble phases
exhibit weaker directional locking 
and reduced or no hysteresis.
The hysteresis occurs in regimes where the particle rearrangements occur
and is most pronounced near the stripe phase.  
We also show that for varied substrate strength,
substrate spacing, and particle
density, 
a number of novel dynamical patterns can form that include
a combination of stripe, bubble, and crystal morphologies. 
\end{abstract}

\maketitle

\section{Introduction}

A wide variety of interacting particle based systems
can couple to periodic one-dimensional (1D) substrates,
including charged colloids
\cite{Chowdhury85,Chakrabarti95,Wei98,Radzihovsky01,Baumgartl04,Zaidouny13},
magnetic skyrmions \cite{Reichhardt16a}, dusty plasmas \cite{Wang18a},
and vortices in nanostructured type-II
superconductors
\cite{Martinoli78,Watson93,Shklovskij06,Dobrovolskiy12,Guillamon14,Dobrovolskiy16,LeThien16,Dobrovolskiy19}.
In these systems, the ordering of the particles depends on
both the filling factor and substrate strength,
and for certain fillings, the system can
form crystal, smectic, or even disordered states.
Under driving, the particles remain trapped until 
a critical depinning force is exceeded.
The magnitude of this critical force
can oscillate as a function of filling,
and for certain fillings, ordered commensurate structures can
appear that are more strongly pinned by the substrate
\cite{Martinoli78,Shklovskij06,LeThien16}.

In general, studies of the driven dynamics of particles on 1D
substrates are performed with the drive applied
perpendicular to the direction of the substrate periodicity.
In some studies, however,
application of driving over a range of different directions reveals
a directional locking effect in which
the particles move along the easy flow direction of the substrate regardless
of the orientation of the applied drive.
When the driving direction is rotated
away from the substrate easy flow direction,
there is a critical angle above which some of the flow can occur
perpendicular to the substrate periodicity direction \cite{Dobrovolskiy19}.
Additionally, most studies of particles
such as colloids or superconducting vortices
on 1D substrates have focused
only on purely repulsive interparticle interactions that produce a
triangular lattice in the absence of a substrate.
In a wide variety of particle based systems, more complicated
particle-particle interactions
are present that lead to the formation
of patterns such as stripes, labyrinths, and bubbles,
where there can be small scale ordering within individual bubbles or stripes as
well as mesoscale ordering of the stripes and bubbles into
lattices
\cite{Seul95,Harrison00,Stoycheva02,Malescio03,Reichhardt03,Nelissen05,Liu08,Edlund10,Reichhardt10,Hooshanginejad24}.
These patterned mesophases can arise when the interparticle interaction
potential has the
short-range attraction and long-range repulsion or SALR form
\cite{Reichhardt03,Reichhardt04,Nelissen05,Reichhardt10,Chen11,McDermott14,Royall18,Liu19,Xu21,Hooshanginejad24}.
Mesophases can also arise if the interparticle interaction potentials
are purely repulsive
but have a multi-step shape or contain multiple length scales
\cite{Malescio03}. In soft matter, patterned mesophases
occur for colloidal systems with various types of
competing interactions \cite{Royall18,Chen11,CostaCampos13,Liu19,Hooshanginejad24}.
In hard condensed matter, mesoscale ordering
can occur for vortices in superconducting systems
with multiple interaction length scales
\cite{Babaev05,Xu11a,Gutierrez12,Komendova13,Curran15,Meng17},
magnetic skyrmion-superconducting vortex 
hybrids \cite{Neto22}, and ordered charges
with competing interactions \cite{Tranquada95,Reichhardt04a}. 

When stripe forming systems with SALR interactions are
placed on a periodic 1D substrate,
the additional periodic modulation of the substrate
allows a variety of new types of aligned stripes and
mixed bubble-stripe states to form \cite{McDermott14}.
In a recent study on the dynamics of a
stripe and bubble forming system driven parallel to the periodicity direction
of a periodic 1D substrate \cite{Reichhardt24},
the stripe states
were strongly pinned since they could easily align with the substrate,
whereas large bubbles were weakly pinned since
they could not fit between adjacent substrate maxima.
Small bubbles were, however, strongly pinned
since they act like point particles
that can fit easily between adjacent substrate maxima.
As a result, the depinning force was
strongly non-monotonic as
a function of the strength of the attractive portion of the SALR potential.
The depinning threshold increased
with increasing attraction in the crystal regime
and reached a local maximum in the stripe state before undergoing
a large drop at the transition to the large bubble state
and then increasing again for small bubbles.
When the substrate was sufficiently weak, the stripes
exhibited elastic depinning
and remained aligned perpendicular to the driving direction.
When the substrate was strong, the stripe depinning was plastic and
the stripes broke apart but, at higher drives,
could dynamically reorder parallel to the drive and perpendicular to the
substrate periodicity direction.
The bubbles could depin either plastically or elastically,
and the transitions between different sliding states for
both the stripe and bubble states were associated with
peaks and dips in the velocity-force curves.

In this work, we place SALR systems
that form anisotropic crystals, stripes, and bubbles
on a 1D periodic substrate
and apply an external drive of fixed magnitude along the easy flow or
$y$ direction, perpendicular to the substrate periodicity direction.
We gradually rotate the drive into the $x$ direction and then
into the $-y$ direction.
A directional locking or guidance effect appears along the easy ($\pm y$)
directions, and there are
a series of dynamical phases that emerge as the driving direction is varied.
For strong substrates, the most pronounced directional locking occurs
for the stripe phase, which exhibits a critical drive angle
$\theta_c$ at which
the particles can jump out of the substrate troughs
through the proliferation of kinks or large-scale plastic deformations.
After the stripes have broken up and plastically deformed,
as the driving angle continues to rotate toward the $x$ direction the
particles 
dynamically reorder into stripes that are oriented
along the $x$ direction,
parallel to the substrate periodicity direction.
As the drive rotates past the $x$ direction and toward the $-y$ direction,
the $x$ orientation of the stripes
persists past $180^\circ-\theta_c$,
leading to strong hysteresis in the directional locking.
This hysteresis is the largest in the stripe regime, and is
strongly reduced or absent in the bubble and anisotropic
crystal phases.
We show that a series of dynamical transitions associated with the
formation of different patterns appear
for the stripe, bubble, and crystal states, and that these transitions
produce signatures in the transport, hysteresis,
and particle ordering.
We map out the dynamic phases for the different patterns
as a function of substrate strength, substrate spacing,
drive amplitude, and filling.

\section{Simulation}

We model a two-dimensional (2D) system with
periodic boundary conditions in the $x-$ and $y$-directions.
The system size is of size
$L$ with $L=36$, and it contains $N$ particles 
at a density of $\rho = N/L^2$. 
The interaction potential combines long-range repulsion and
short-range attraction. Previously, it was shown that this potential
can produce crystals, stripes, void lattices, and bubble lattices
depending on the relative strength of the attractive and repulsive
terms as well as the particle density
\cite{Reichhardt03, Reichhardt04,Reichhardt10,Reichhardt24}.
The dynamics of particle $i$ are obtained by integrating the following overdamped equation of motion:
\begin{equation}
\eta \frac{d {\bf R}_{i}}{dt} =
-\sum^{N}_{j \neq i} \nabla V(R_{ij}) + {\bf F}^{s}_{i} +
        {\bf F}_{D} ,
\end{equation}
where $\eta=1$ is the damping term.
The particle-particle  interaction potential is
\begin{equation}
V(R_{ij}) = \frac{1}{R_{ij}} - B\exp(-\kappa R_{ij}),
\end{equation}
where the location of particle $i (j)$ is
${\bf R}_{i (j)}$ and 
$R_{ij}=|{\bf R}_i-{\bf R}_j|$.

The first term in Eq.~(2)
is a repulsive Coulomb interaction. It dominates at
long distances and also at short distances, so the particles cannot
all collapse to a point.
The second term is a short-range attraction of
strength $B$ that falls off exponentially with a screening length of $\kappa$.
For fixed $\rho$ and $\kappa$, as $B$ increases the system
forms crystal, stripe, and clump states,
while for fixed $B$ and $\kappa$,
as $\rho$ increases, the system forms clumps, stripes,
void lattices, and a high-density uniform crystal
\cite{Reichhardt10,Reichhardt24}.
In this work, we fix $\kappa = 1.0$ while varying $B$ and $\rho$.

The second term in Eq.~(2) is the substrate force,
modeled as a sinusoidal potential,
\begin{align}
{\bf F}_s^i = F_p \cos(2\pi x_i/a_p)
\end{align} where $x_i$ is the $x$ position of particle $i$
and the maximum substrate force is $F_{p}$.
The substrate is composed of $N_p$ minima and
has a lattice constant of $a_{p} = L/N_{p}$.

After we initialize the particle positions, we
apply a driving force given by
\begin{equation}
{\bf F}_{D}= F_D [\cos(\theta){\hat {\bf y}} + \sin(\theta){\hat {\bf x}}] \ .
\end{equation}
Initially, $\theta=0^\circ$ and the drive is in the $+y$
direction. We then increase $\theta$ in increments to a value
of $180^\circ$, such that when $\theta=90^\circ$ the drive is aligned with
the $+x$ direction, and at the end of our sweep the driving
is in the $-y$ direction.
Throughout this work we fix $F_{D}= 1.25$.
We also consider a rotating drive in which $\theta$ is increased
in increments from $0^\circ$ to $360^\circ$ repeatedly.
As the driving angle is varied, we measure
$\langle V\rangle = \sum^{N}_i{\bf v}_i\cdot {\hat {\bf x}}$ and
$\langle V\rangle = \sum^{N}_i{\bf v}_i\cdot {\hat {\bf y}}$.
In general,
we find that $\langle V_{y}\rangle$
exhibits a sinusoidal signature that is in phase with the drive angle,
but $\langle V_{x}\rangle$ shows stronger variations
due to the coupling to the substrate.

\begin{figure}
  \centering
  \includegraphics[width=\columnwidth]{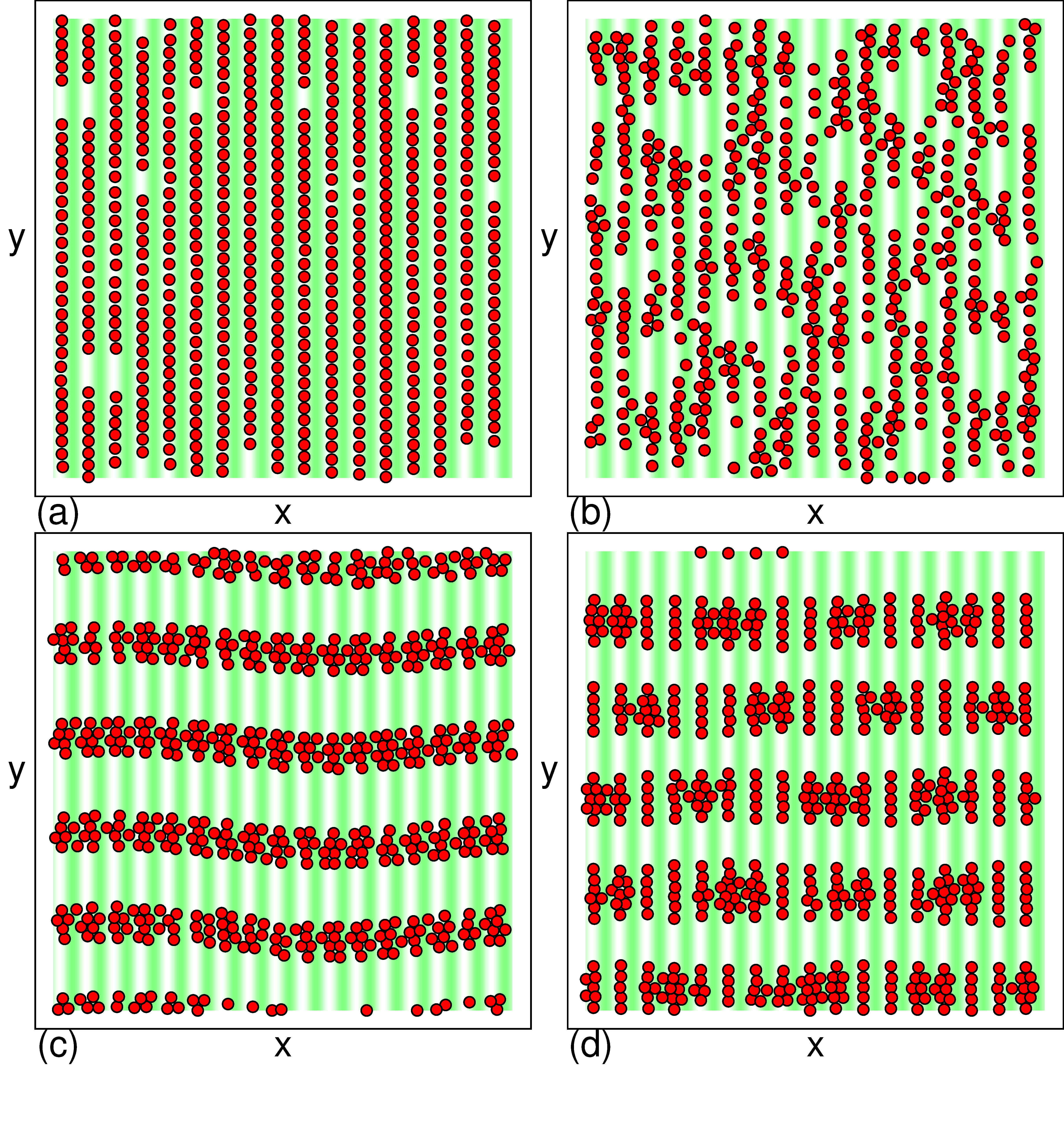}
\caption{Particle positions (red circles) and substrate maxima (green) and
  minima (white) for a system with
$B = 2.25$, $F_{p} = 1.25$, a substrate
  spacing of $a_{p}= 2.067$, and $\rho = 0.454$.
  The drive is initially applied in the $+y$ direction and rotated
  through the $+x$ direction to the $-y$ direction.
(a) A parallel stripe state at $\theta = 0^\circ$ with no drive.
(b) A disordered state
at $\theta=66.5^\circ$, just above the drive angle for which the particles
begin to move along the $x$ direction.
(c) A perpendicular stripe state at $\theta = 90^\circ$
where the driving is purely along the $x$ direction.
(d) Persistence of the stripe-like structure at $\theta = 160^\circ$.
} 
\label{fig:1}
\end{figure}

\begin{figure}
  \centering
  \includegraphics[width=\columnwidth]{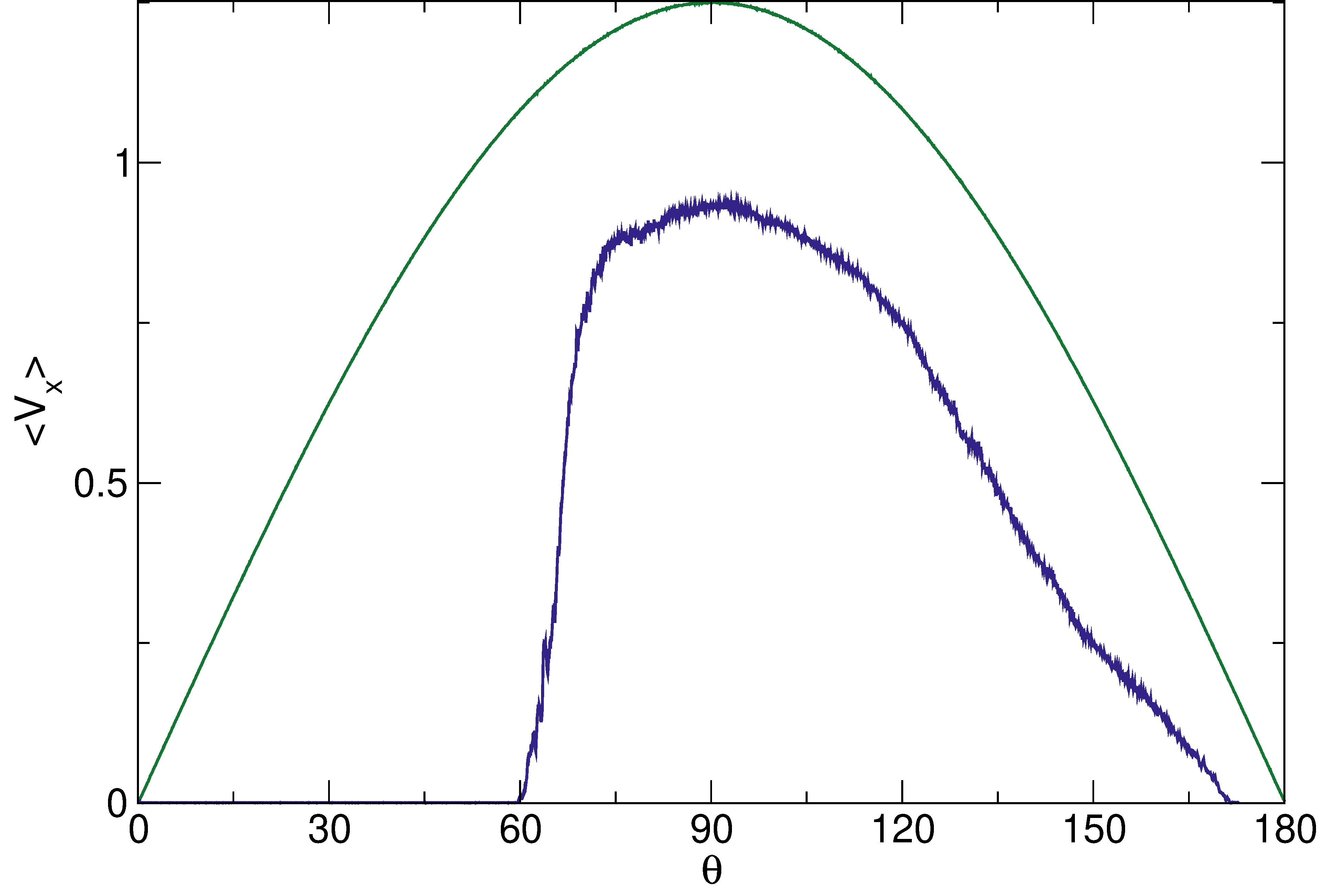}
\caption{$\langle V_{x}\rangle$ vs drive angle $\theta$
  at $B=2.25$, $a_p=2.067$, and $\rho=0.454$ for the
  system from Fig.~\ref{fig:1} with $F_p=1.25$ (blue)
  and $F_{p} = 0.0$ (green).
  When $F_{p} = 0.0$, the velocity response is
  symmetric about $\theta=90^\circ$,
  while for $F_{p} = 1.25$ the velocity is strongly asymmetric,
  with the motion remaining locked to the $y$ direction
  over a much larger range of angles for the $0^\circ <\theta<90^\circ$
  window than for the $90^\circ<\theta<180^\circ$ window.	}
 \label{fig:2}
\end{figure}

\section{Results}

In Fig.~\ref{fig:1}, we show images from
a system with $B = 2.25$, substrate spacing $a_{p}= 2.067$, and
density $\rho = 0.454$. 
In the absence of a substrate, stripes form
that are each two or three particles in width.
When the substrate is present but no drive is applied,
pinned stripes appear that are a single particle wide,
as shown in Fig.~\ref{fig:1}(a).
There are also several void-like features
due to the attractive part of the SALR interaction potential.
For drive orientations in the range  $0 < \theta < 60^\circ$,
the system has the same appearance as that shown
in Fig.~\ref{fig:1}(a), but the particles are sliding
along the $+y$ direction.
In Fig.~\ref{fig:2}
we plot $\langle V_{x}\rangle$ versus $\theta$ for the same system with
no substrate, $F_p=0.0$, and for a substrate with $F_p=1.25$.
In the presence of the substrate,
$\langle V_{x}\rangle = 0.0$ when $\theta < 60^\circ$,
indicating that the motion is locked
along the $y$-direction. The particles have the same structure
shown in Fig.~\ref{fig:1}(a).
For $\theta > 60^\circ$, $\langle V_{x}\rangle$ increases rapidly and
reaches a plateau near $\theta = 75^\circ$ followed by
a local maximum at $\theta=90^\circ$.
When $\theta>90^\circ$, $\langle V_x\rangle$ decreases but
does not reach zero until
$\theta = 173^\circ$, showing that there is a strong asymmetry in
the velocity response between the $0^\circ < \theta < 90^\circ$
window and the $90^\circ < \theta < 180^\circ$ window.
If the response had been symmetric,
$\langle V_{x}\rangle$ would have reached zero near
$\theta = 110^\circ$.
This asymmetry arises because the particles form
different patterns during different stages of the drive orientation.
We note that the $\langle V_{y}\rangle$ response is symmetric,
with $\langle V_y\rangle=1.25$ at $\theta=0^\circ$ decreasing to
$\langle V_y\rangle=-1.25$ at $\theta=180^\circ$. This is the same response that
would
be expected for motion in the absence of pinning,
since the 1D substrate only affects $x$ direction motion; thus,
$\langle V_y\rangle$ has no interesting features for
all the cases we consider in this work.

\begin{figure}
  \centering
  \includegraphics[width=\columnwidth]{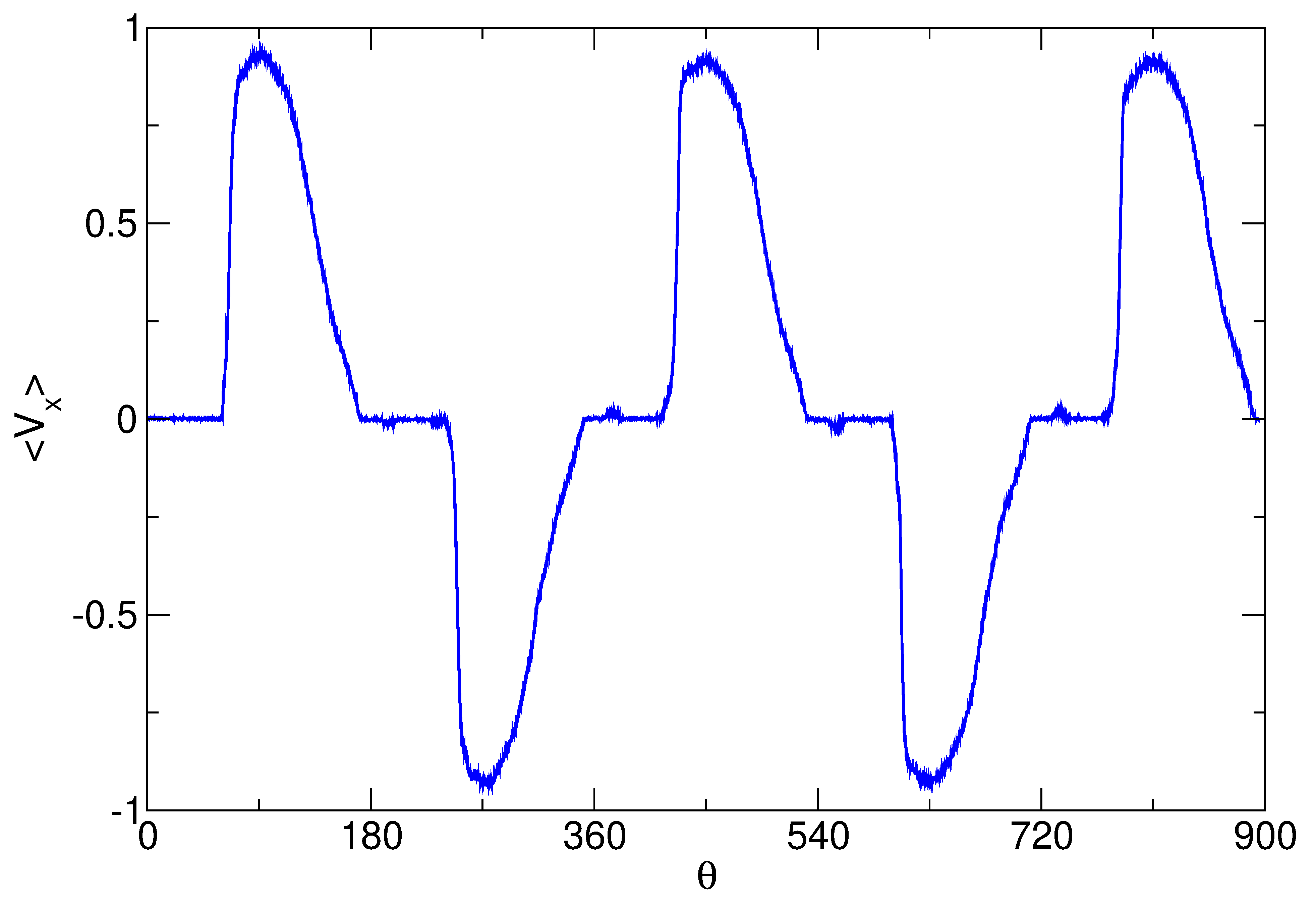}
\caption{$\langle V_{x}\rangle$ vs drive angle $\theta$ for the system
  from Fig.~\ref{fig:1} with $B=2.25$, $F_p=1.25$, $a_p=2.067$, and
  $\rho=0.454$
  for two and a half complete rotations of the drive,
  showing that the asymmetry in the velocity response persists
over multiple drive cycles.
}
\label{fig:3}
\end{figure}

In Fig.~\ref{fig:1}(b), we illustrate the particle positions at
$\theta = 66.5^\circ$, just above the drive angle at which
$\langle V_{x}\rangle$ becomes finite.
The system is
strongly disordered, and the motion of some particles remains
locked along the $y$ direction while other particles
move in both the $y$ and $x$ directions, 
so that a disordered plastic flow state appears.
At $\theta = 75^\circ$, the system dynamically orders into an
$x$-oriented stripe state that correlates with the rapid increase
in $\langle V_x\rangle$.
The $x$ oriented stripes are pictured in
Fig.~\ref{fig:1}(c) at $\theta = 90^\circ$, and
as the drive is rotated toward the $-y$-direction
this stripe pattern persists over a range of angles that is wider than
the range for which it appeared when the drive was still partially
oriented in the $+y$-direction.
As the drive continues to rotate toward the $-y$ direction,
the general shape of the stripe remains relatively constant,
but an increasing number of particles
become pinned in the $x$ direction and have their motion locked along
the $-y$ direction,
as shown in Fig.~\ref{fig:1}(d) at $\theta = 160^\circ$.
For $\theta = 180^\circ$, the configuration
looks the same as the
$\theta=0^\circ$ configuration in Fig.~\ref{fig:1}(a),
but the particles are now moving in the $-y$ direction.
If we continue to rotate the drive to values larger than
$\theta=180^\circ$, the same set of patterns and asymmetric
responses are repeated, 
as shown in Fig.~\ref{fig:3} for the system
from Fig.~\ref{fig:1} undergoing two and a half
complete rotations of the drive. The response remains asymmetric for
each drive cycle.

\begin{figure}
  \centering
  \includegraphics[width=\columnwidth]{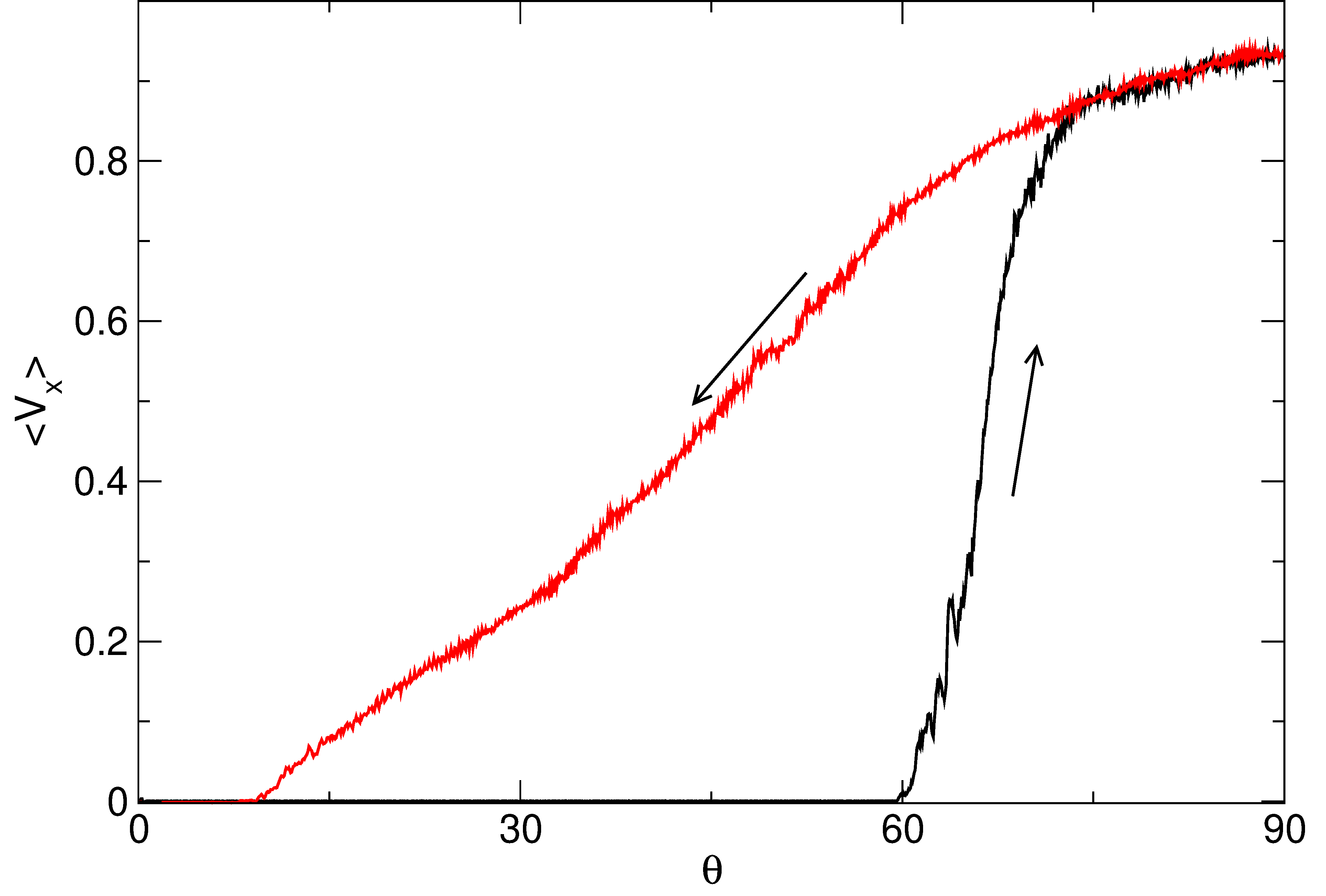}
\caption{Hysteretic velocity response
  for the system
  from Fig.~\ref{fig:1} with $B=2.25$, $F_{p} = 1.25$, $a_p=2.067$,
  and $\rho=0.454$.
  Black: 
  $\langle V_{x}\rangle$ vs $\theta$ for increasing
  $\theta$ from $0^\circ$ to $90^\circ$.
  Red: Inverted response
  $\langle V_{x}\rangle$ vs $90^\circ-\theta$ for $\theta$ increasing
  from $90^\circ$ to $180^\circ$. Inverting the large $\theta$ portion of
  the curve illustrates the hysteresis more clearly.}
        \label{fig:4}
\end{figure}

To show more clearly the asymmetry of the hysteresis,
in Fig.~\ref{fig:4} we plot
$\langle V_x\rangle$ versus 
$\theta$ for $\theta$ increasing from $\theta=0^\circ$ to
$\theta=90^\circ$, and overlay an inverted plot of
$\langle V_x\rangle$ versus $90^\circ-\theta$ for $\theta$ increasing from
$\theta=90^\circ$ to
$\theta=180^\circ$.
If the response had been the same on each half of the
drive rotation, the curves would follow each other.
Instead, we find a strong 
hysteresis effect, and $\langle V_x\rangle$ is much larger for
$\theta>90^\circ$ than for $\theta<90^\circ$
due to the persistence
of the $x$-oriented stripe pattern shown in Fig.~\ref{fig:1}(c,d).

\begin{figure}
  \centering
  \includegraphics[width=\columnwidth]{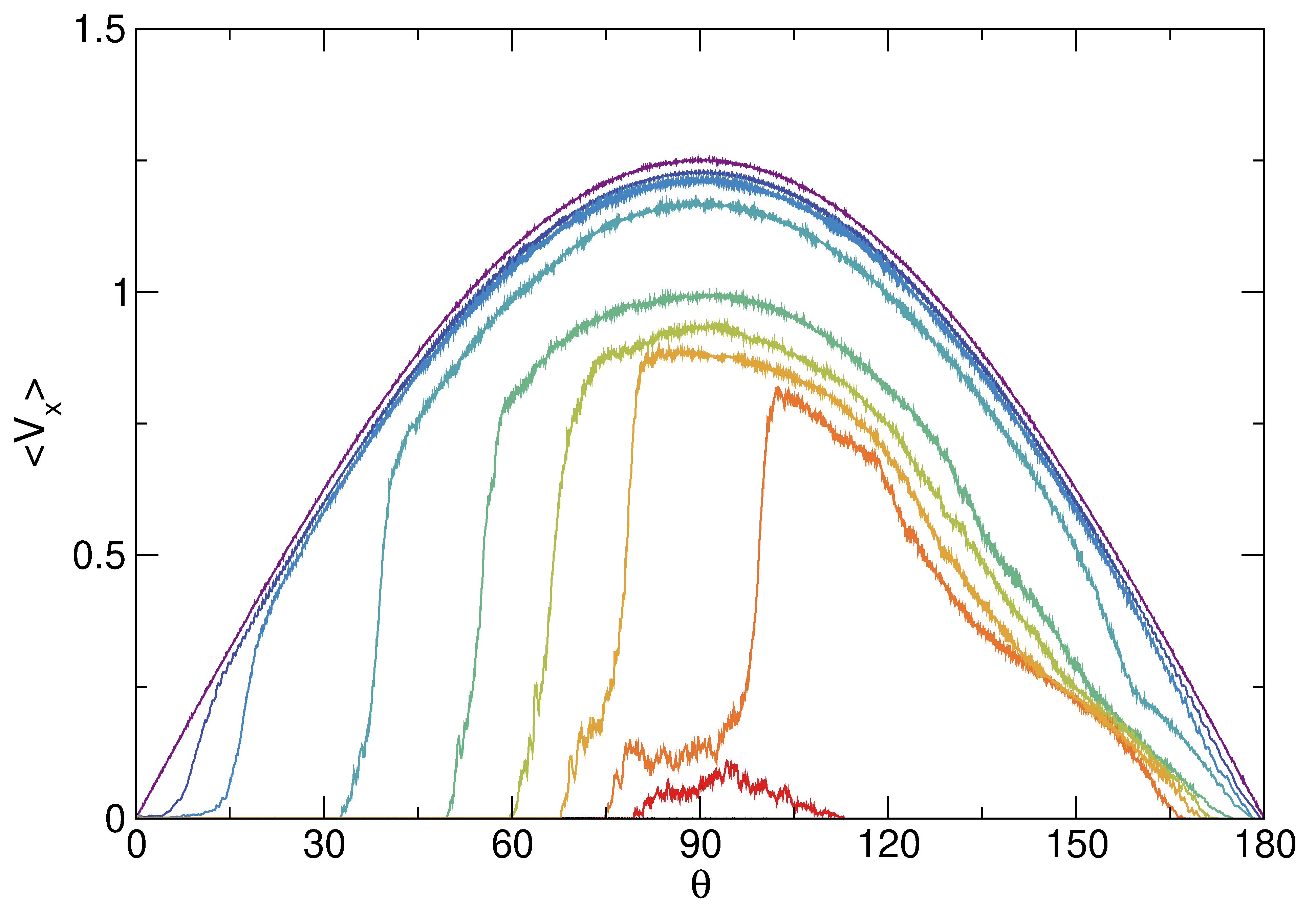}
\caption{$\langle V_{x}\rangle$ vs $\theta$
  for the system from Fig.~\ref{fig:1} with $B=2.25$, $a_p=2.067$, and
  $\rho=0.454$ at
  $F_{p} = 0.0$, 0.5, 0.672, 0.875, 1.125, 1.25, 1.325, 1.375, 1.4, and $1.5$,
  from top to bottom.
  For $F_{p} > 1.45$, the motion is locked
  along the $\pm y$ direction for all $\theta$.
}
\label{fig:5}
\end{figure}

We next consider the evolution of the asymmetry for varied
substrate strengths.
In Fig.~\ref{fig:5}, we plot
$\langle V_{x}\rangle$ versus $\theta$ for the system
from Fig.~\ref{fig:1} at
$F_{p} = 0.0$, 0.5, 0.672, 0.875, 1.125, 1.25, 1.325, 1.375, 1.4, and $1.5$.
For $F_{p} > 1.45$, the motion remains locked
along $\pm y$ for all drive angles.
When $F_{p} = 1.4$, there is a small hopping motion along
$x$ at drives near $\theta=90^\circ$.
The curves are much more symmetric
for $F_{p} < 0.7$, while at $F_{p} = 1.375$,
the velocity response is strongly asymmetric.

\begin{figure}
  \centering
  \includegraphics[width=\columnwidth]{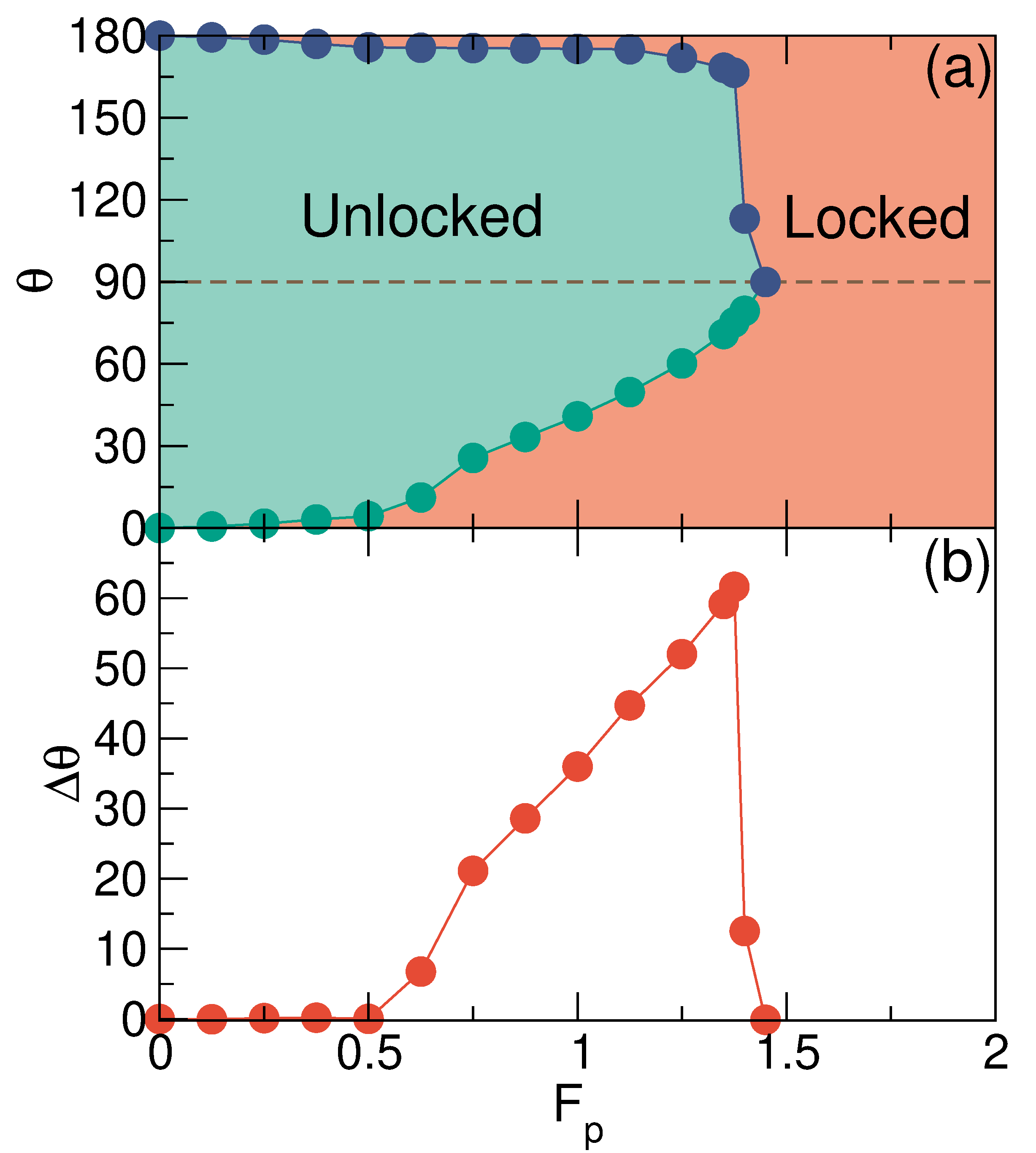}
\caption{(a) Boundaries $\theta_{\rm low}$ (green)
  and $\theta_{\rm hi}$ (blue) separating the regions where the motion is locked and not locked along the $\pm y$ direction plotted as a function of $\theta$ vs $F_p$ for a sample with $B=2.25$, $a_p=2.067$, and $\rho=0.454$.
  The locked motion is along $+y$ below the dashed line and along $-y$
  above the dashed line.
(b) $\Delta \theta = \theta_{\rm low} - (180 -\theta_{\rm hi})$ vs $F_{p}$ for
  the same system.
Larger values of $\Delta \theta$ correspond to a greater amount of
hysteresis.
} 
\label{fig:6} 
\end{figure}

\begin{figure}
  \centering
  \includegraphics[width=\columnwidth]{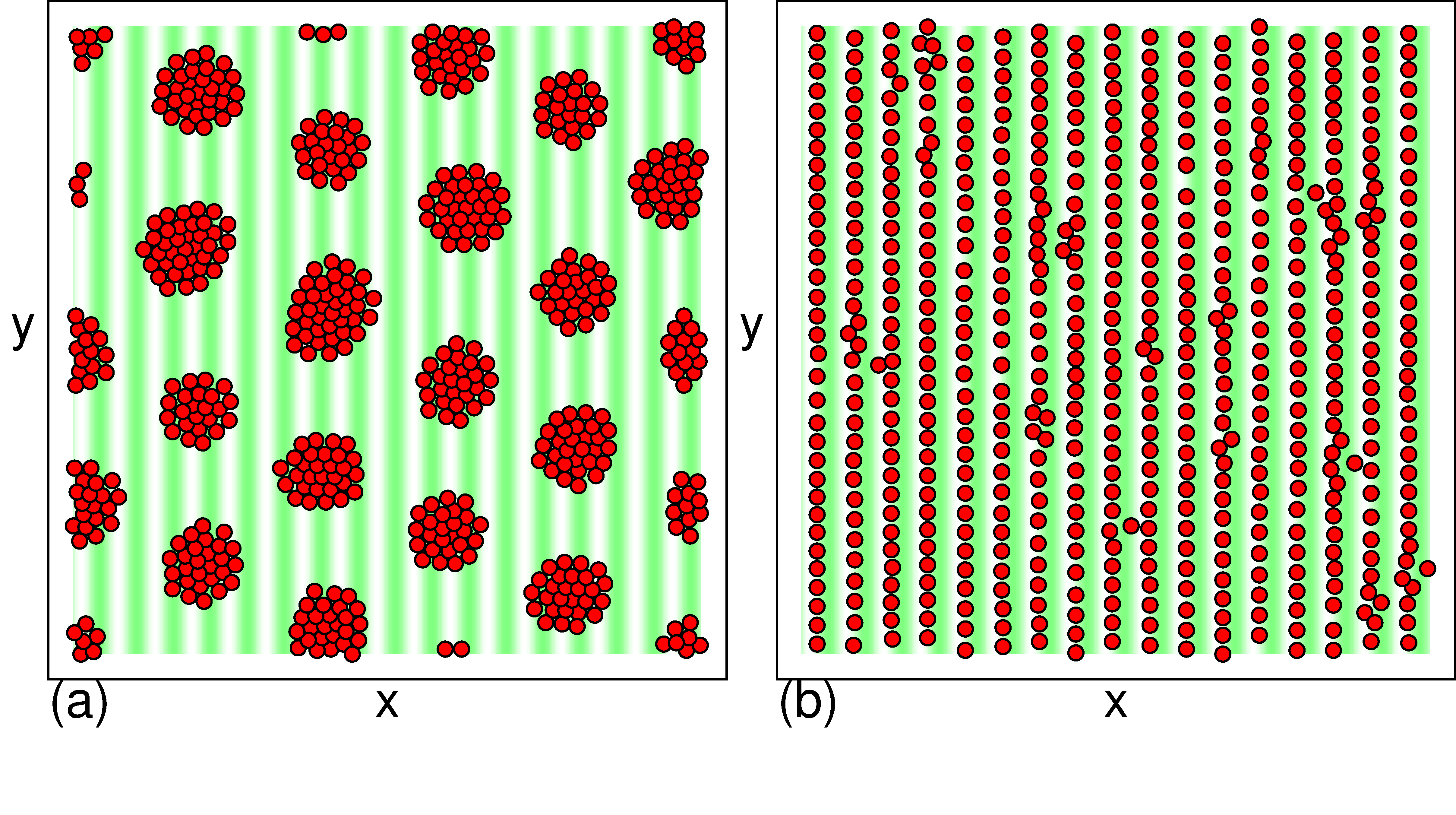}
\caption{Particle positions (red circles) and substrate maxima (green) and
minima (white) for the system from Fig.~\ref{fig:6}
with $B=2.25$, $a_p=2.067$, and $\rho=0.454$.
(a) The bubble phase at $F_p = 0.5$ and $\theta = 30^\circ$.
(b) The locked stripe phase
at $F_{p} = 1.4$ and $\theta = 90^\circ$.
A small amount of hopping along the $x$ direction occurs
in this state.
}       
\label{fig:7} 
\end{figure}

To quantify the evolution of the regions
over which the flow is locked along  $\pm y$
for increasing $F_{p}$, in Fig.~\ref{fig:6}(a) we plot
as a function of $\theta$ versus $F_p$
the lower
boundary $\theta_{\rm low}$ and upper boundary $\theta_{\rm hi}$
separating values of $\theta$ where
$\langle V_{x}\rangle= 0.0$ and
the motion is locked 
from those where
$\langle V_x\rangle$ is finite and
the motion is unlocked.
The width of the lower locked region increases rapidly
for $F_{p} > 0.5$, and when
$F_p > 1.45$, the unlocked region disappears and the motion is always
locked to the $\pm y$ direction.
We highlight the asymmetry of the locking response
in Fig.~\ref{fig:5}(b),
where we plot the difference in width of the $+y$ and $-y$ locking
regimes, $\Delta\theta = \theta_{\rm low} - (180^\circ -\theta_{\rm hi})$, versus
$F_p$.
If the widths are symmetric, this measure will be zero and there will be
no hysteresis.
We find that for $F_p<0.6$,
$\Delta\theta=0$, while
for higher $F_{p}$, $\Delta \theta$
increases linearly with increasing $F_{p}$
until it reaches a maximum at $F_{p} = 1.375$ and
then rapidly decreases back to zero at $F_{p} = 1.45$.
For $F_{p} < 0.6$, instead of forming a stripe, the system forms
bubbles, as shown
in Fig.~\ref{fig:7}(a) for $\theta = 30^\circ$ and $F_{p} = 0.5$.
At large  $F_{p}$, the system can no longer form
$x$-direction oriented stripes, and the hysteresis disappears.
An example is illustrated in 
Fig.~\ref{fig:7}(b)
at $F_{p} = 1.4$ and $\theta = 90^\circ$,
where, although some
individual particle hopping in $x$ does occur,
the particles are mostly pinned along the $x$ direction
and there
are no $x$-direction oriented stripes.

\begin{figure}
  \centering
  \includegraphics[width=\columnwidth]{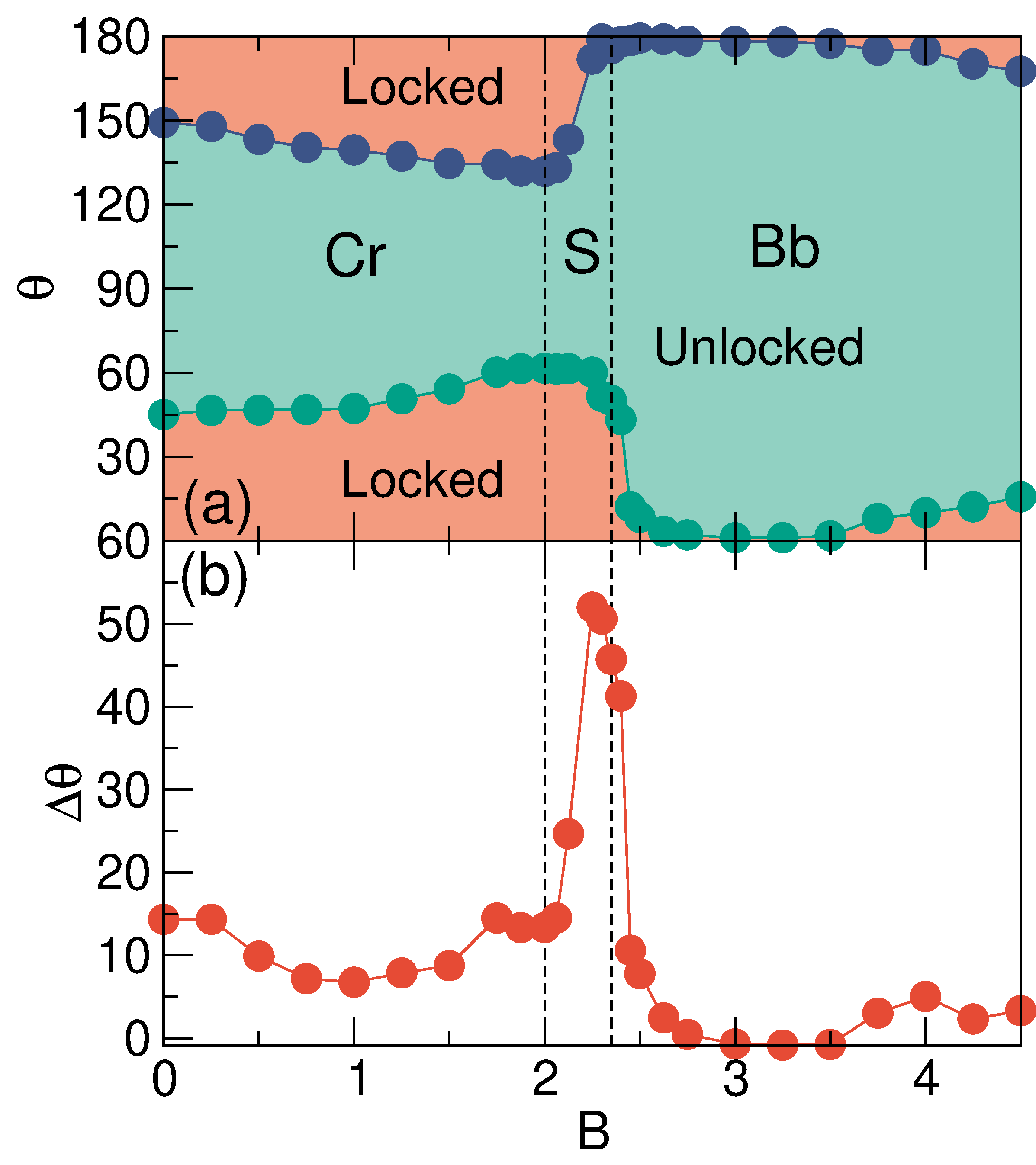}
\caption{(a) Boundaries $\theta_{\rm low}$ (green) and $\theta_{\rm hi}$ (blue)
  separating the regions where the motion is locked and not locked along the
  $\pm y$ direction plotted as a function of $\theta$ vs $B$ for
  the system from Fig.~\ref{fig:1} with $F_p=1.25$, $a_p=2.067$,
  and $\rho=0.454$.
  Dashed lines indicate the values of $B$ at which the system forms
  crystal (Cr), stripe (S), and bubble (Bb) phases.
  (b) The corresponding $\Delta \theta$ vs $B$.
  The hysteresis is largest in the stripe forming regime.
}
   \label{fig:8}
\end{figure}

By varying the strength $B$ of the attractive term, we
next show that the hysteresis is the most produced in the stripe phase.
In the absence of a substrate, the SALR particles
form an anisotropic crystal for $0 < B < 1.9$,
stripes for $2.0 < B < 2.35$, and bubbles for $ B> 2.35$.
In Fig.~\ref{fig:8}(a), we plot the boundaries $\theta_{\rm low}$ and
$\theta_{\rm high}$ separating the regions where flow is locked and
unlocked in the $y$ direction as a function of $\theta$ versus $B$ for
a system with $F_p=1.25$.
As $B$ increases, the width of the locked regimes increase,
reaching a maximum for $B = 2.25$ at the transition into the stripe state
before decreasing nearly to zero. Windows of finite $y$-direction locking
reappear deep in the bubble state for
$B > 2.4$.
Figure~\ref{fig:8}(b) shows the corresponding $\Delta \theta$
versus $B$, which peaks strongly 
in the stripe phase
near $B = 2.25$,
not far from the transition to a bubble state.
This indicates that by far the most pronounced hysteresis arises in the
stripe state.
There is still a modest amount of hysteresis
in the crystal phase but almost no hysteresis
in
the bubble phase for $2.5 < B < 3.5$.
When $B > 3.5$, the bubbles begin to
shrink in size due to the increasing strength of the attractive term,
leading to the appearance of some plasticity in the depinning
transition and an increase in the hysteresis for the largest
values of $B$ considered here.

\begin{figure}
  \centering
  \includegraphics[width=\columnwidth]{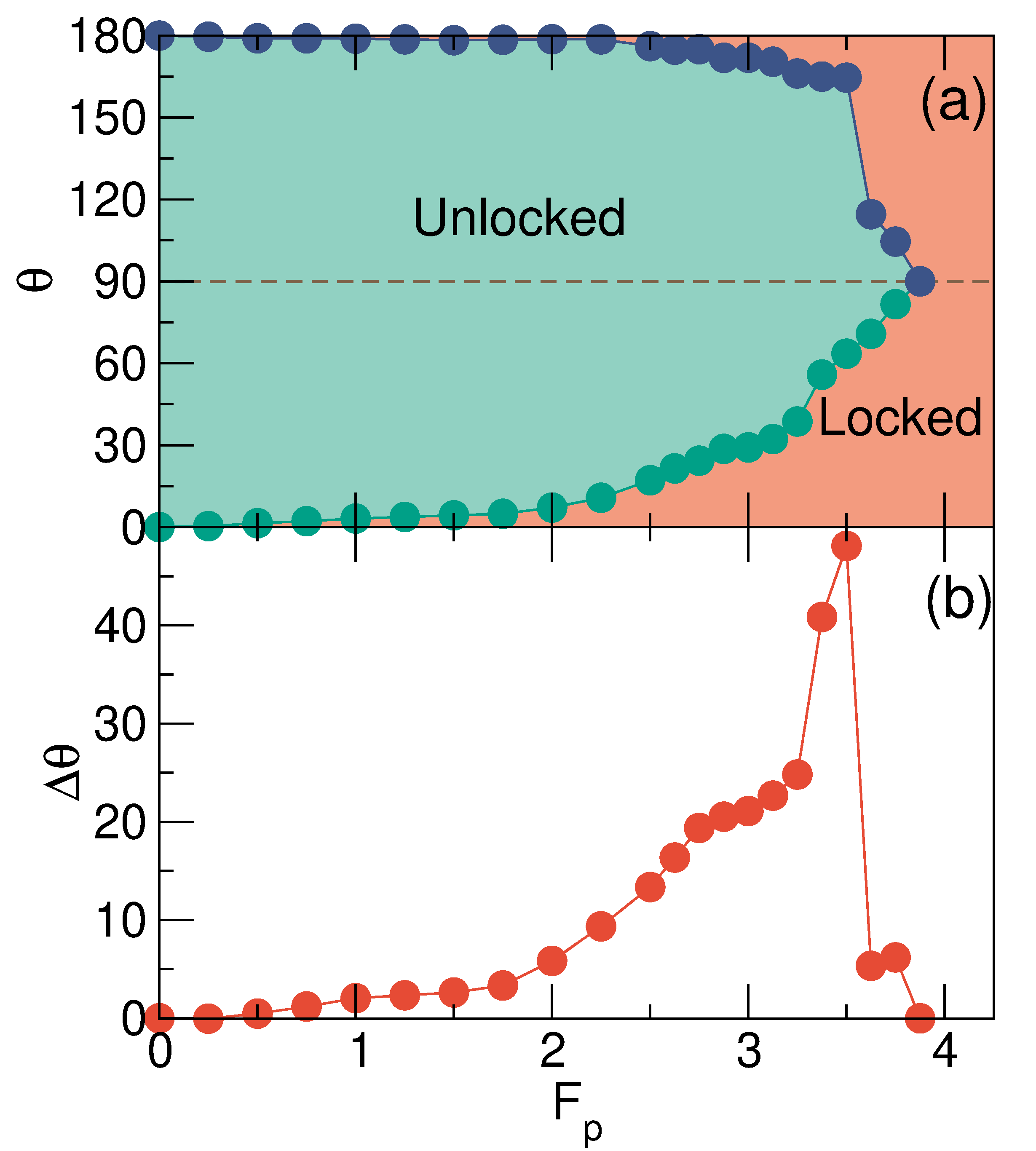}
\caption{(a) Boundaries $\theta_{\rm low}$ (green) and $\theta_{\rm hi}$ (blue)
  separating the regions where the motion is locked and not locked along
  the $\pm y$ direction plotted as a function of $\theta$ vs $F_p$ for the
  system from Fig.~\ref{fig:1} in the bubble regime with
  $B=2.6$, $a_p=2.067$, and $\rho=0.454$.
  (b) The corresponding $\Delta \theta$ vs $F_{p}$.
}
        \label{fig:9}
\end{figure}

We next focus on hysteresis in the bubble state.
If $F_p$ is large enough, 
the bubbles break apart into stripes that are
oriented in the $x$ direction, and when this happens,
hysteresis can occur.
In the stripe phase for $B = 2.25$, the motion becomes
completely locked to the $y$ direction for $F_{p} > 1.45$, while
in the bubble state at $B = 2.6$,
complete $y$-direction locking occurs only when $F_{p} > 3.7$.
Figure~\ref{fig:9}(a) shows the boundaries
$\theta_{\rm low}$ and $\theta_{\rm hi}$ separating the regions in which
the motion is locked and not locked to the $\pm y$ direction plotted
as a function of $\theta$ versus $F_p$ for the system from Fig.~\ref{fig:1}
in the bubble state at $B=2.6$.
In Fig.~\ref{fig:9}(b), the corresponding $\Delta \theta$ versus $F_p$
is small for $F_{p} < 1.5$,
where the system remains in a well ordered bubble state for all
values of $\theta$. The hysteresis
grows with increasing $F_{p}$ and the system breaks up into a stripe state
for $F_p>2.0$. A maximum in $\Delta \theta$ appears at
$F_{p} = 3.5$, and $\Delta \theta$ drops to zero
for $F_{p} > 3.75$.

\begin{figure}
  \centering
\includegraphics[width=\columnwidth]{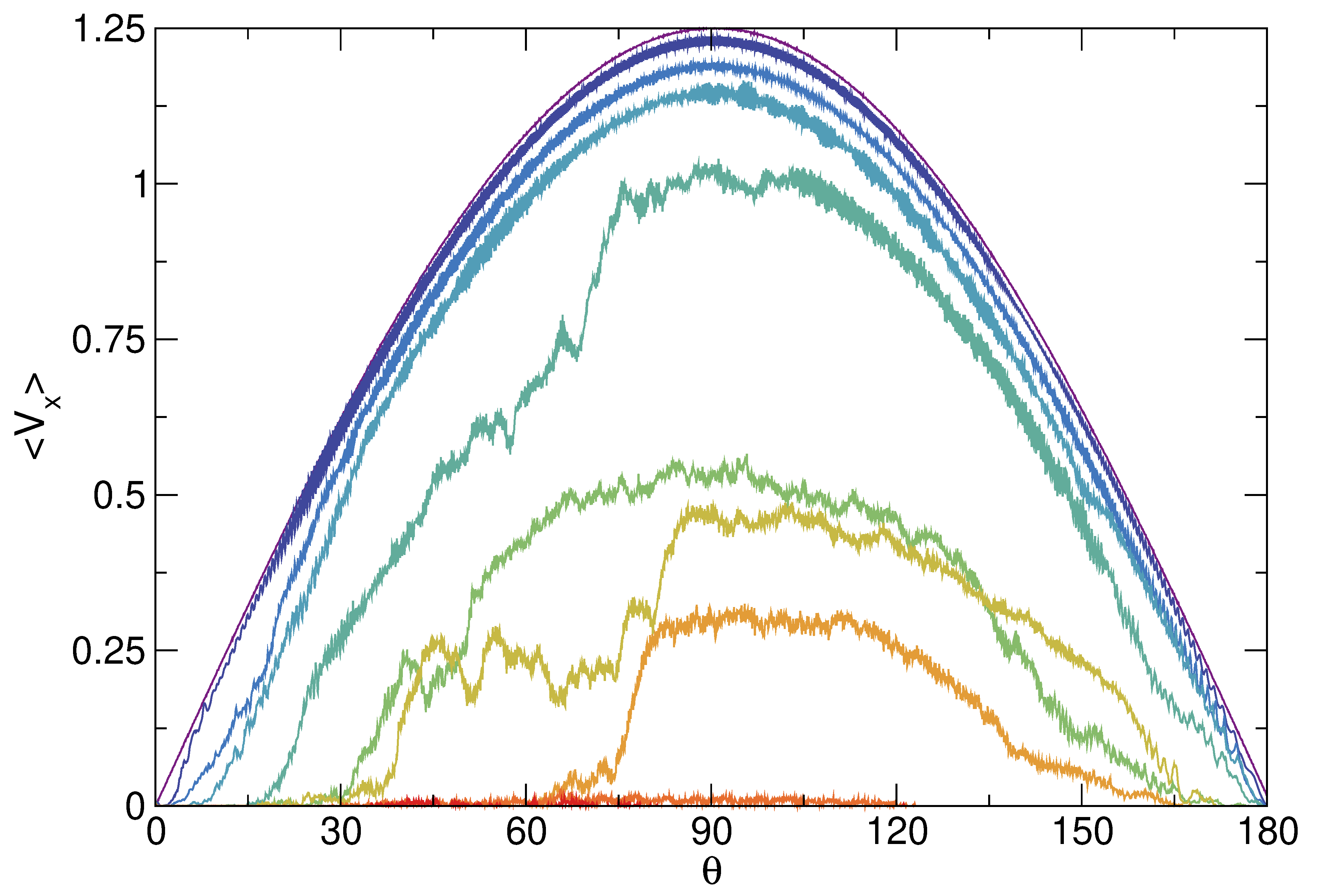}
\caption{ $\langle V_{x}\rangle$ vs $\theta$ for the bubble
  forming system from
  Fig.~\ref{fig:9} with $B=2.6$, $a_p=2.067$, and $\rho=0.454$ at
  $F_{p} = 0.0$, 0.75, 1.5, 2.0, 2.5, 3.0, 3.125, 3.5, 3.625, and $3.75$,
  from top to bottom.
}
\label{fig:10}
\end{figure}

\begin{figure}
  \centering
  \includegraphics[width=\columnwidth]{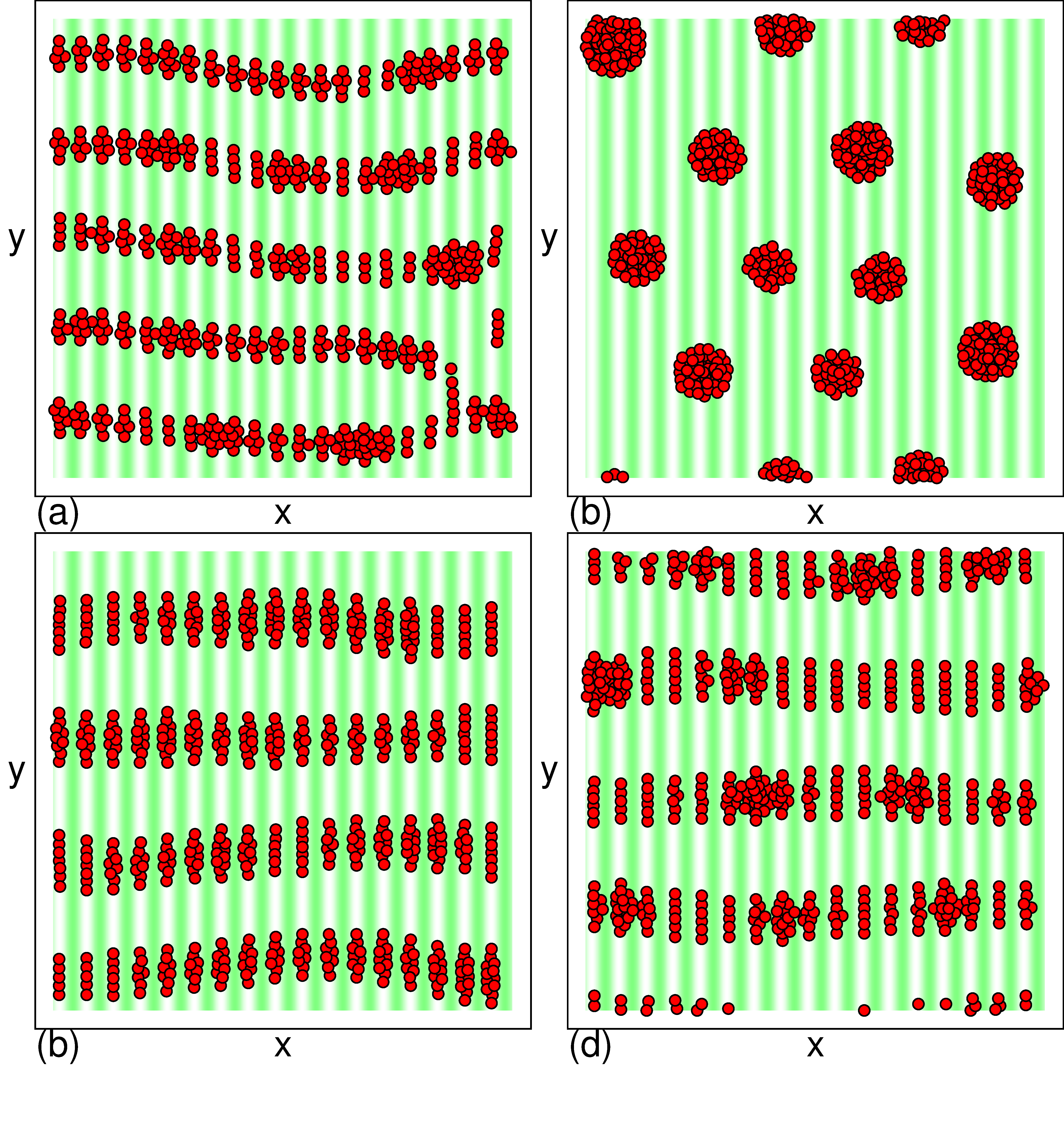}
\caption{Particle positions (red circles) and substrate maxima (green)
  and minima (white) for the system from Fig.~\ref{fig:10} with
$B = 2.6$, $a_p=2.067$, and $\rho=0.454$.
(a) $F_{p}= 1.5$ and $\theta = 10^\circ$.
(b) $F_{p} = 1.5$ and $\theta = 35^\circ$.
(c) $F_{p} = 3.5$ and $\theta = 0.0^\circ$.
(d) $F_{p}= 3.5$ and $\theta = 90^\circ$.
}
\label{fig:11}
\end{figure}

In Fig.~\ref{fig:10} we
plot $\langle V_{x}\rangle$ versus $\theta$ for the
$B=2.6$ system from Fig.~\ref{fig:9} at
$F_{p} = 0.0$, 0.75, 1.5, 2.0, 2.5, 3.0, 3.125, 3.5, 3.625, and $3.75$.
The system forms a bubble state at $F_p = 1.25$.
The curves are relatively symmetric about $\theta=90^\circ$
for $F_{p} < 1.75$, while when $F_{p} > 1.75$, the
curves  become increasingly
asymmetric, which correlates with the growth in
$\Delta \theta$ shown
in Fig.~9(b).
For $F_{p} \geq 1.5$, 
the system forms a $y$-direction locked 
stripe of the type shown in Fig.~\ref{fig:11}(a)
at $F_p=1.5$ and $\theta = 10.0^\circ$.
As the drive rotates to higher $\theta$,
this stripe state transitions into a bubble state,
as shown in Fig.~\ref{fig:11}(b) for
$F_p=1.5$ and $\theta = 35^\circ$.
For $F_{p} > 1.75$,
the system forms a $y$-direction locked state consisting of
$x$-direction aligned stripes that only occupy the
substrate troughs,
as shown in Fig.~\ref{fig:11}(c) at
$F_p=3.5$ and $\theta = 0.0^\circ$.
As the drive rotates, a portion of the
particles begin to  move along the $x$ direction
by assembling into bubbles that travel along the stripes,
as illustrated in Fig.~\ref{fig:11}(d) at
$F_{p} = 3.5$ and $\theta = 90^\circ$.

\begin{figure}
  \centering
  \includegraphics[width=\columnwidth]{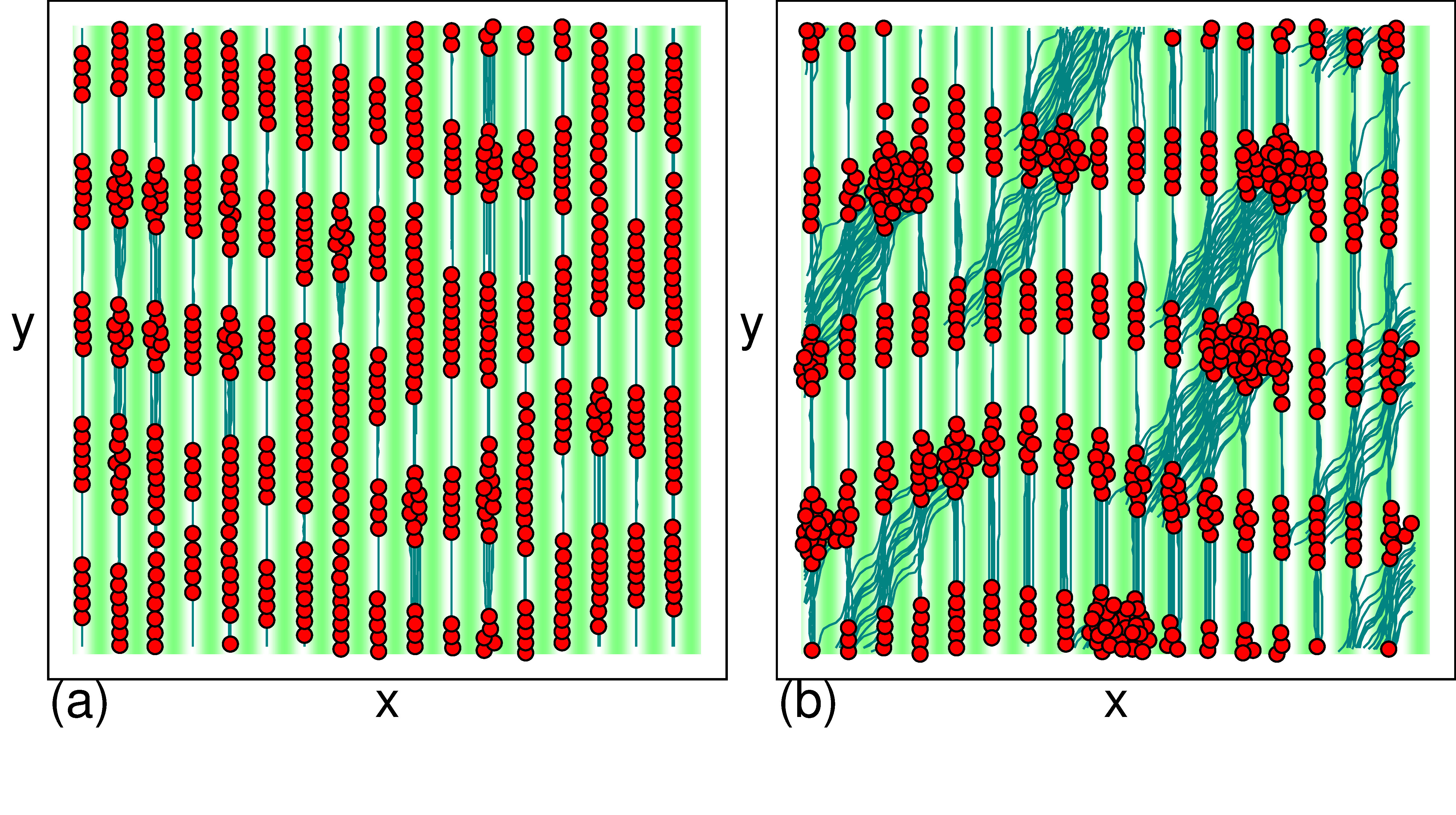}
\caption{Particle positions (red circles), substrate maxima (green) and
  minima (white), and particle trajectories (blue lines) for
  the system from Fig.~\ref{fig:10} with $B = 2.6$,
  $a_p=2.067$, $\rho=0.454$, and $F_{p}= 3.125$.
  (a) The locked phase at $\theta = 0.0^\circ$
  where the motion is only along the  $+y$ direction.
  (b) The stripe-bubble phase at $\theta=30^\circ$, where particles 
  in the stripe segments move only along $+y$
  while particles in the localized bubbles are moving along
  both $+x$ and $+y$.
}
        \label{fig:12}
\end{figure}

To more clearly visualize the dynamics in the
$B=2.6$ and $F_{p} > 1.75$
stripe-bubble
coexistence phase 
that is responsible for the hysteresis peak
from Fig.~\ref{fig:9}(b),
in Fig.~\ref{fig:12}(a) we plot
the particle trajectories for a fixed period of time at
$F_{p} = 3.125$ in the $y$-direction locked regime
at $\theta = 0.0^\circ$.  Here the particles form a
stripe pattern that occupies only the substrate troughs,
and the motion is strictly locked to the $+y$ direction.
Figure~\ref{fig:12}(b) shows the same system at $\theta = 30^\circ$,
where particles in the striped portions of the system move
only along the $+y$ direction,
while particles that are inside the localized bubbles
move along both the $+x$ and $+y$ directions.

\begin{figure}
  \centering
  \includegraphics[width=\columnwidth]{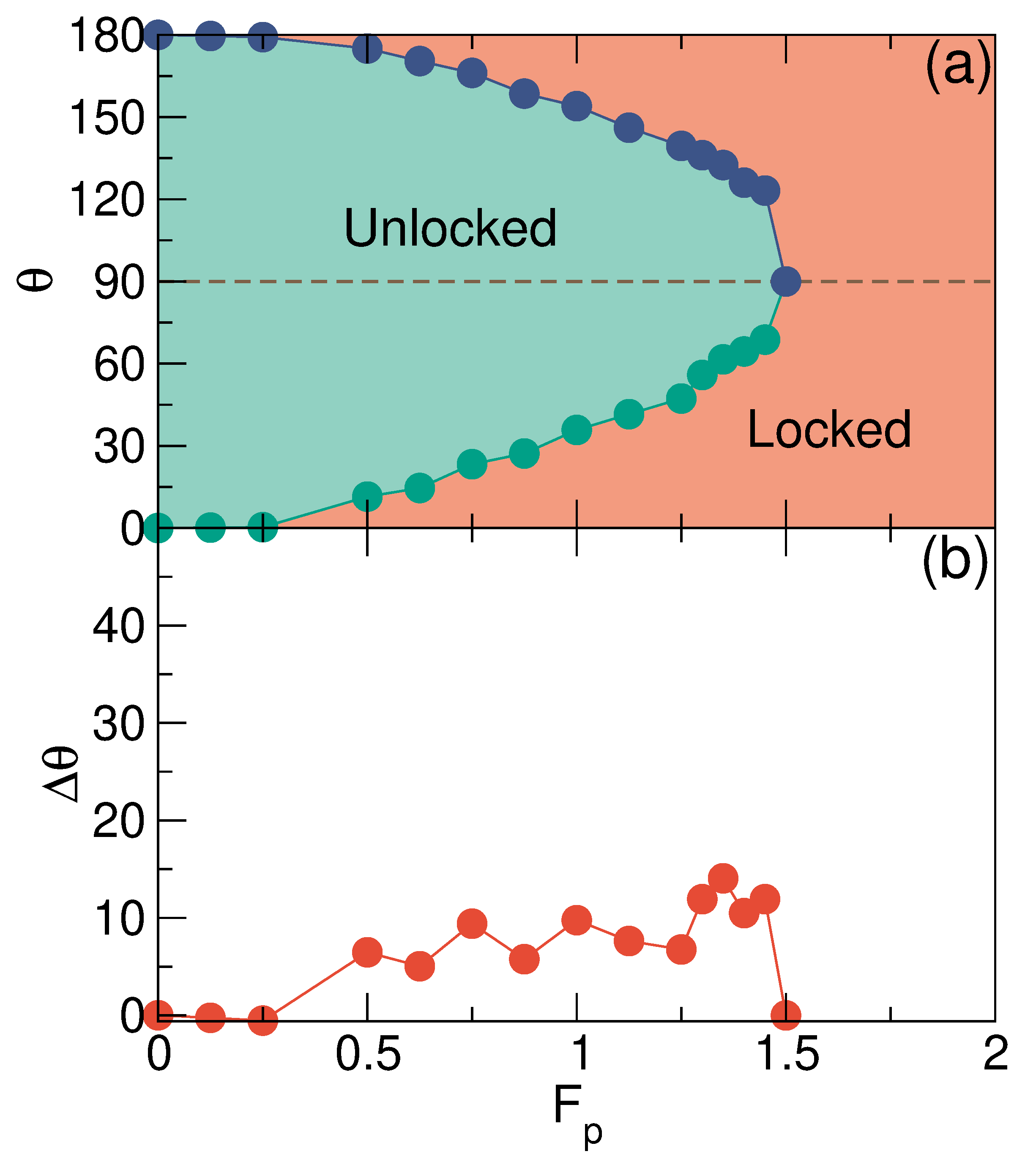}
\caption{Boundaries $\theta_{\rm low}$ (green) and $\theta_{\rm hi}$ (blue)
separating the regions where the motion is locked and not locked along
the $\pm y$ direction plotted as a function of $\theta$ vs $F_p$
for the system from Fig.~\ref{fig:8} with $a_p=2.067$ and
$\rho=0.454$ at $B=1.0$ in the anisotropic crystal regime.
(b) The corresponding $\Delta \theta$ vs $F_{p}$,
showing strongly reduced hysteresis.
}
\label{fig:13}
\end{figure}

\begin{figure}
  \centering
  \includegraphics[width=\columnwidth]{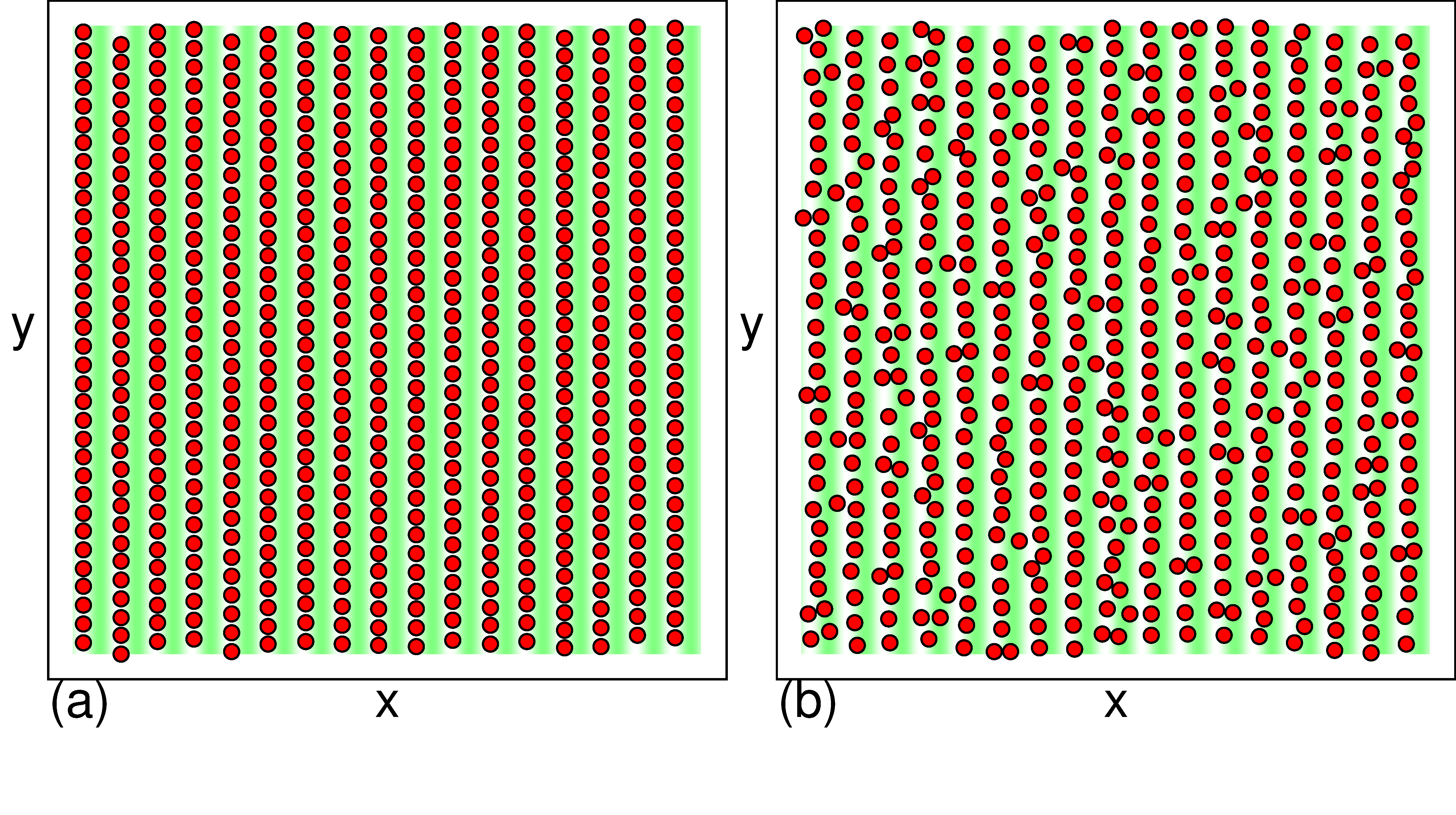}
\caption{Particle positions (red circles) and substrate maxima (green) and
minima (white) for the system from Fig.~\ref{fig:13} with $a_p=2.067$
and $\rho=0.454$
in the anisotropic crystal regime at $B=1.0$ and $F_p=1.35$.
(a) At $\theta = 0.0^\circ$, there is an anisotropic crystal.
(b) At $\theta = 90^\circ$, the
particles are disordered.
}
    \label{fig:14}
\end{figure}

In Fig.~\ref{fig:13}(a) we plot the boundary separating the regions
of locked and unlocked $y$-direction motion as a function of $\theta$
versus $F_p$ for the system from Fig.~\ref{fig:8} at $B=1.0$ in the
anisotropic crystal regime, and in Fig.~\ref{fig:13}(b) we show the
corresponding $\Delta \theta$ versus $F_p$.
The motion becomes locked to the $y$-direction for all values of $\theta$
once 
$F_{p} > 1.475$, a threshold that is
slightly higher than what we observed for the stripe phase.
As indicated by Fig.~\ref{fig:13}(b),
the hysteresis is considerably smaller than
what appears in the stripe or bubble phases.
In Fig.~\ref{fig:14}(a) we show the particle positions
in the $y$-direction locked phase at $\theta = 0.0^\circ$ for the system
from Fig.~\ref{fig:13} at  $F_{p} = 1.35$,
where the particles form a uniform anisotropic crystal.
At $\theta=90^\circ$ in the same system,
Fig.~\ref{fig:14}(b) indicates that a disordered state
with a uniform density appears.
In general, the hysteresis is reduced throughout the
anisotropic crystal phases
because no dramatic
structural transitions
occur as the drive is rotated,
unlike the large rearrangements that appear
for the stripe and bubble states.

\begin{figure}
  \centering
  \includegraphics[width=\columnwidth]{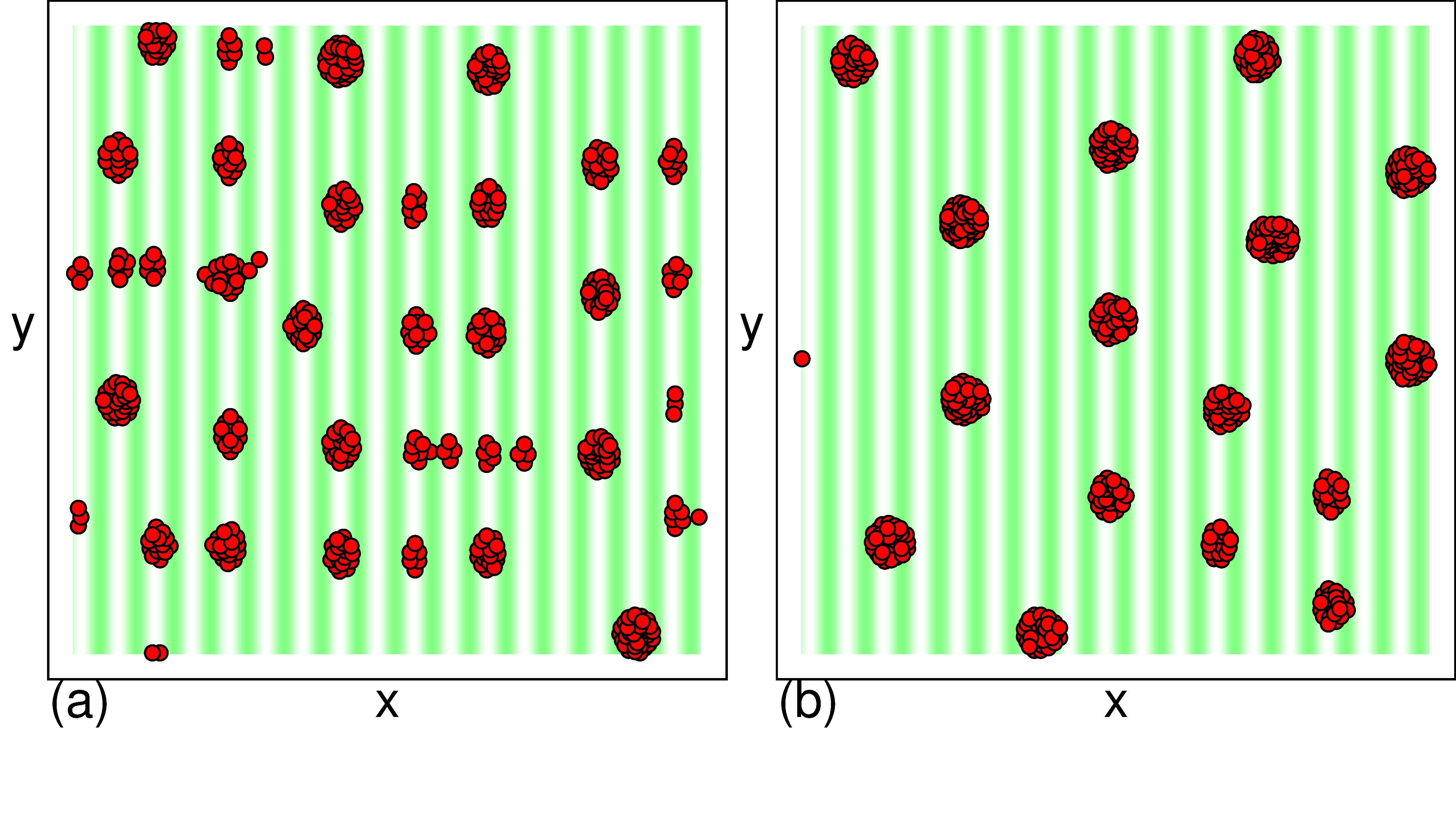}
\caption{Particle positions (red circles) and substrate maxima (green) and
  minima (white) for the system from Fig.~\ref{fig:8} with $a_p=2.067$ and
  $\rho=0.454$ at
$B = 4.0$  and $F_{p} =  4.0$.
  (a) A mixture of stripe segments and bubbles at
  $\theta = 0.0^\circ$. (b) A bubble state at $\theta = 90^\circ$.
	}
        \label{fig:15}
\end{figure}

We observe reduced hysteresis for other values of $B$ in the crystal phase.
In the bubble phase, we find hysteresis
only when $F_{p}$ is large enough to partially break apart
some of the bubbles.
In Fig.~\ref{fig:15}(a), we show the particle
positions at $B  = 4.0$ and $F_p = 4.0$
in the $y$-direction locked state at $\theta = 0.0^\circ$,
where a combination of stripe segments and bubbles are present.
In the same system at
$\theta = 90^\circ$, shown in Fig.~\ref{fig:15}(b),
an unlocked bubble phase appears. Once the bubbles have assembled, they
remain stable out to values of $\theta$ that are further from $\theta=90^\circ$,
leading to hysteresis in the locking response.

\begin{figure}
  \centering
  \includegraphics[width=\columnwidth]{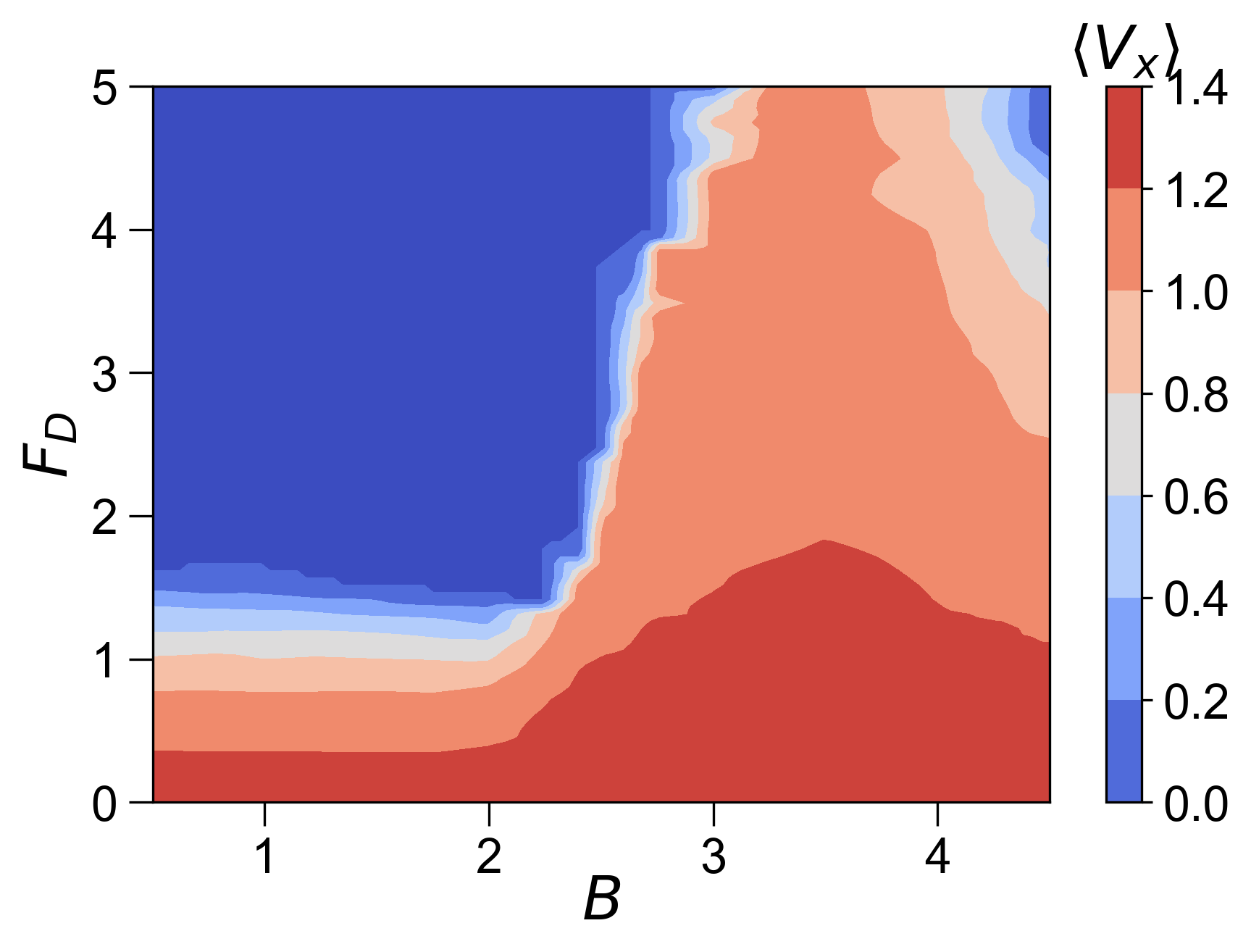}
\caption{Heat map of $\langle V_x\rangle$ as a function of
$F_D$ vs $B$ for a system with $a_p=2.067$, $\rho=0.454$, and  
$\theta = 90^\circ$.
The system passes through crystal, stripe, large bubble,
and small bubble states with increasing $B$.
}
\label{fig:16}
\end{figure}

In Fig.~\ref{fig:16}, we plot a heat map of
$\langle V_x\rangle$ as a function of $F_D$ versus $B$
for a system with $\theta = 90^\circ$.
As $B$ increases, the system passes through
crystal, stripe, large bubble, and smaller bubble states.
The complete locking of the motion
along the $y$ direction
is the most pronounced in the crystal and stripe
phases for $F_{p} > 1.5$.
In the large bubble regime for $3.0 < B < 3.9$,
no $y$ direction locking occurs up to at least
$F_{p}=5.0$.
At larger $B$, there is a partial reentrance
in the $y$ direction locking when the bubbles decrease
in size and become better pinned by the substrate troughs,
since the smaller bubbles act more like point particles than
the larger bubbles do.

\section{Varied Particle Density}

\begin{figure}
  \centering
  \includegraphics[width=\columnwidth]{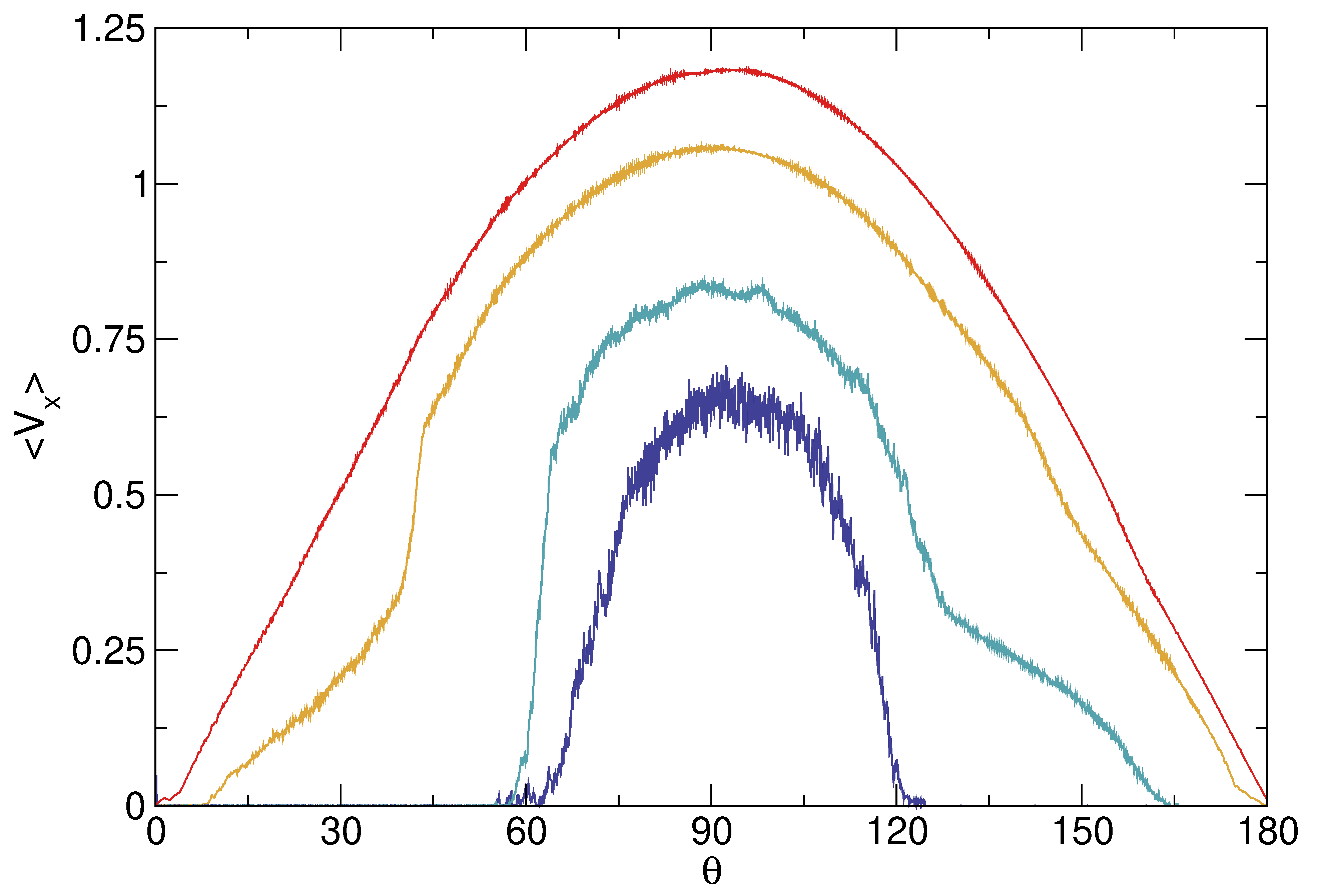}
\caption{$\langle V_{x}\rangle$ vs $\theta$ for
a system with $B=2.25$, $F_p=1.25$, and $a_p=2.067$ at
 $\rho = 0.164$, 0.33, 0.682,  and $1.32$, from bottom to top. 
}
        \label{fig:17}
\end{figure}

We next consider the effect of varying the particle density $\rho$
for the system from Fig.~\ref{fig:1} with $B = 2.25$ and $F_{p} = 1.25$.
In Fig.~\ref{fig:17} we plot $\langle V_{x}\rangle$ versus
$\theta$ at $\rho = 0.164$, 0.33, 0.682,  and $1.32$.
When $\rho = 0.164$,
there are large windows of $\theta$ over which
$\langle V_{x}\rangle= 0.0$, and the curve is nearly
completely symmetric about $\theta=90^\circ$.
In contrast, the response is strongly asymmetric
for $\rho = 0.33$, which has an extended 
finite $\langle V_x\rangle$ velocity tail for $\theta > 90^\circ$.
The $\rho=0.682$ curve is more symmetric, and has a clear jump up in
$\langle V_x\rangle$ near $\theta = 35^\circ$.
For $\rho = 1.32$, the particle-particle interactions begin to overwhelm
the substrate interactions, and the curve becomes nearly symmetric with
no $y$-direction locking windows.

\begin{figure}
  \centering
  \includegraphics[width=\columnwidth]{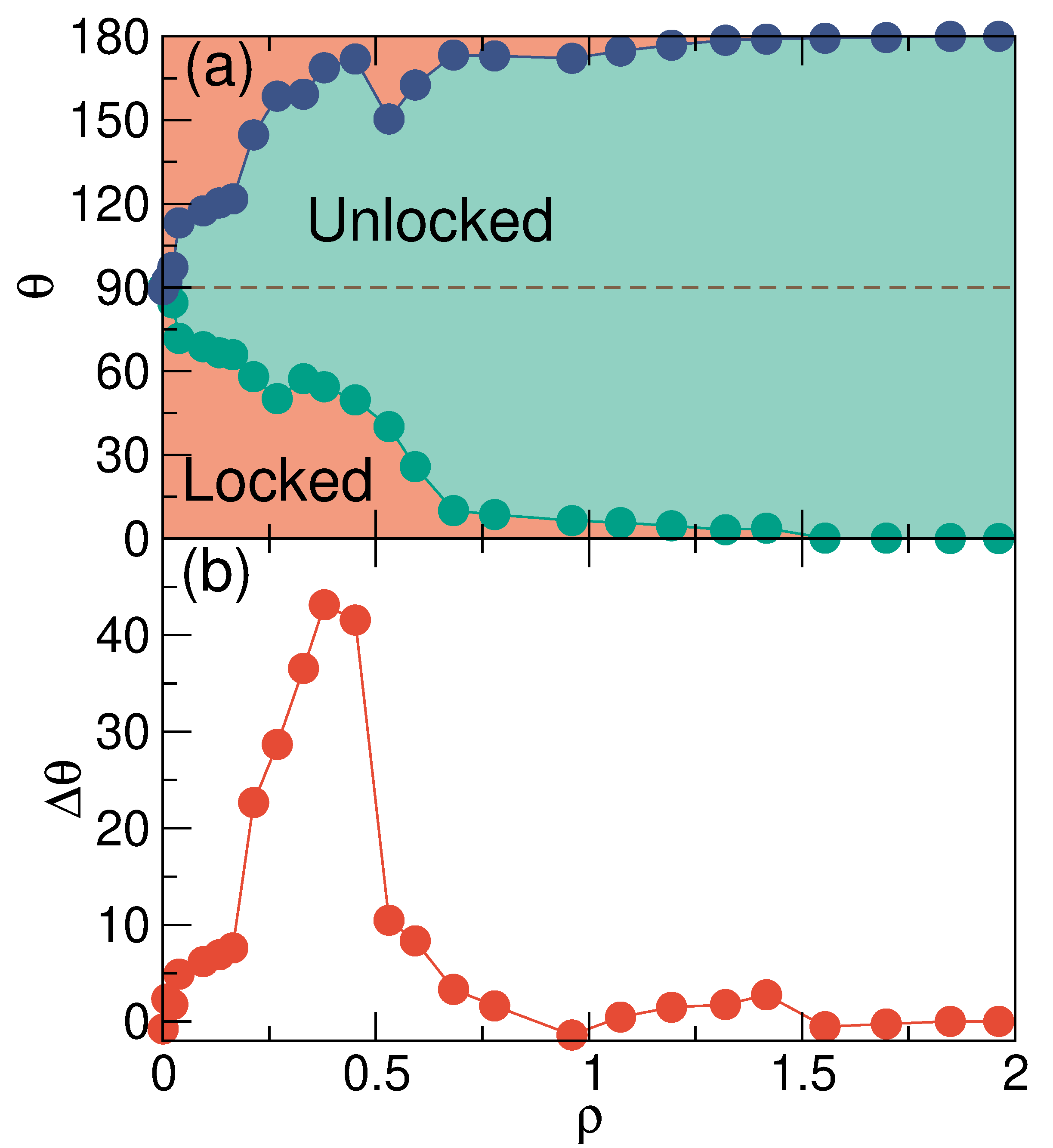}
\caption{Boundaries $\theta_{\rm low}$ (green) and $\theta_{\rm hi}$ (blue)
  separating the regions where the motion is locked and not locked along
  the $\pm y$ direction plotted as a function of $\theta$
  vs $\rho$
for the
system from Fig.~\ref{fig:17} with $B=2.25$, $a_p=2.067$, and $F_{p} = 1.25$. 
(b) The corresponding $\Delta \theta$ vs $\rho$.
}
 \label{fig:18}
\end{figure}

In Fig.~\ref{fig:18}(a) we plot the
boundaries separating the locked and unlocked $y$-direction
motion regimes as a function of
$\theta$ versus $F_{p}$ in the system from Fig.~\ref{fig:17}.
Figure~\ref{fig:18}(b) shows
the corresponding $\Delta \theta$ versus $F_{p}$.
At low particle densities, thin stripes or small bubbles are present,
giving a window of
reduced hysteresis.
For $0.19 < \rho < 0.52$, we find
strong hysteresis when the 
system
forms an aligned stripe phase as the driving
direction approaches $90^\circ$.
For high densities, a stripe-bubble coexistence phase is present.
In general, when $F_{p} = 1.25$, the
pining is strongly reduced at the higher densities,
so the amount of hysteresis is reduced even though structural
transitions may still occur
as the drive is rotated.

\begin{figure}
  \centering
\includegraphics[width=\columnwidth]{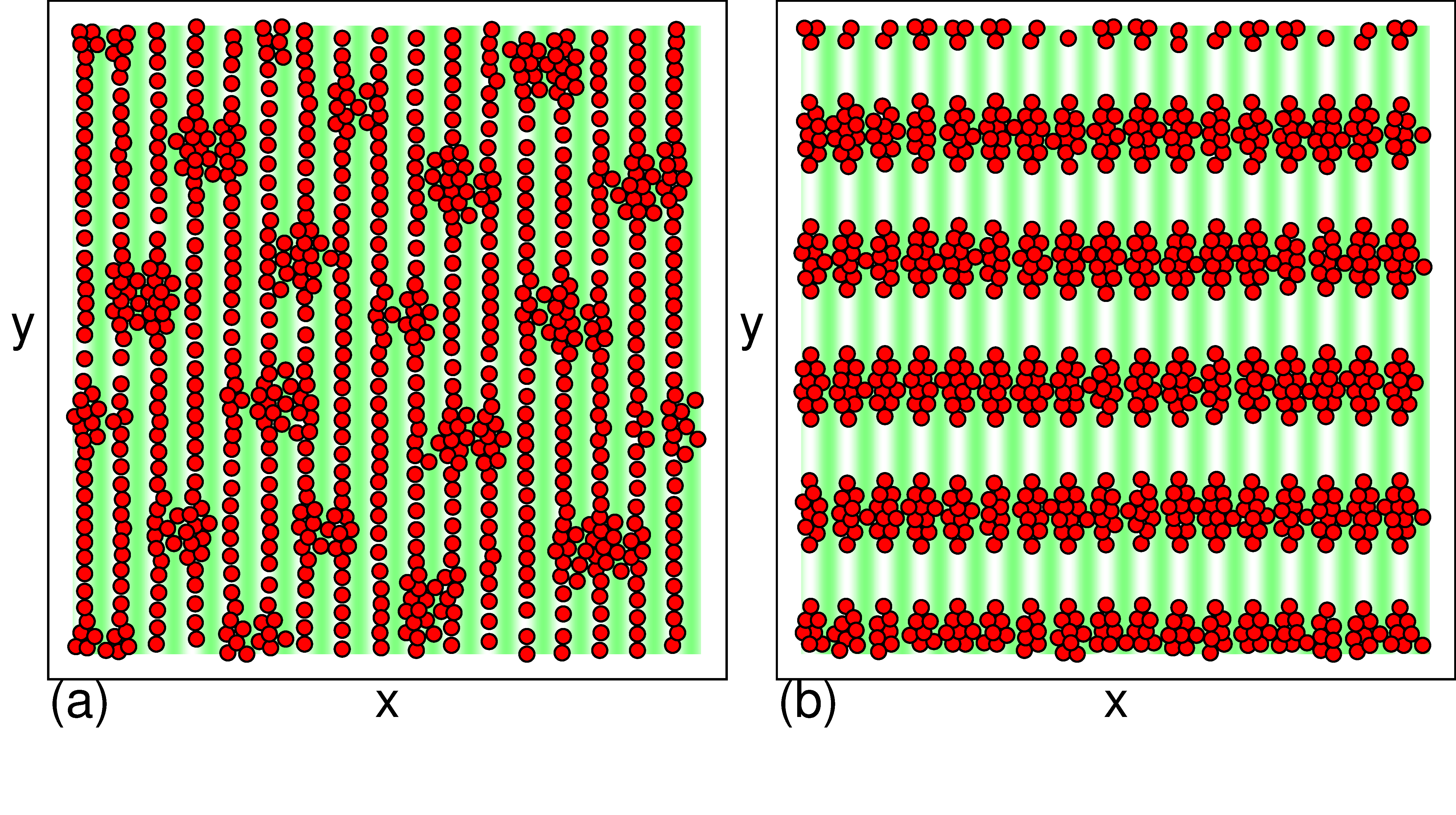}
\caption{Particle positions (red circles) and substrate maxima (green) and
  minima (white) for the system from Fig.~\ref{fig:17} with $a_p=2.067$,
  $B=2.25$, $F_p=1.25$, and $\rho=0.68$.
(a) The stripe-bubble phase at $\theta = 20^\circ$.
(b) The stripe phase at $\theta = 90^\circ$.
}
        \label{fig:19}
\end{figure}

In Fig.~\ref{fig:19}(a), we show the coexisting stripe-bubble state
from the system in Fig.~\ref{fig:17} at
$\rho = 0.68$ and $\theta = 20^\circ$.
The motion of the particles in the striped portion of the system is
locked along the $+y$ direction, while the particles in the bubbles move
along both the $+x$ and $+y$ directions.
At $\theta=90^\circ$, the system forms stripes that are well-aligned in
the $x$ direction and that are composed of bubbles that are elongated
in the $y$ direction,
as shown in Fig.~\ref{fig:19}(b).
The jump up in $\langle V_x\rangle$
that occurs near $\theta = 30^\circ$ in
Fig.~\ref{fig:17} for the $\rho=0.68$ sample
coincides with a transition from the
stripe-bubble state to the stripe state.
In this case, simply measuring hysteresis according to $\Delta \theta$,
which is based on the locking of the velocity to the $\pm y$ direction,
gives a small hysteresis value; however, inspection of the actual
finite value of $\langle V_x\rangle$ reveals that
strong velocity hysteresis is present.

\begin{figure}
  \centering
\includegraphics[width=\columnwidth]{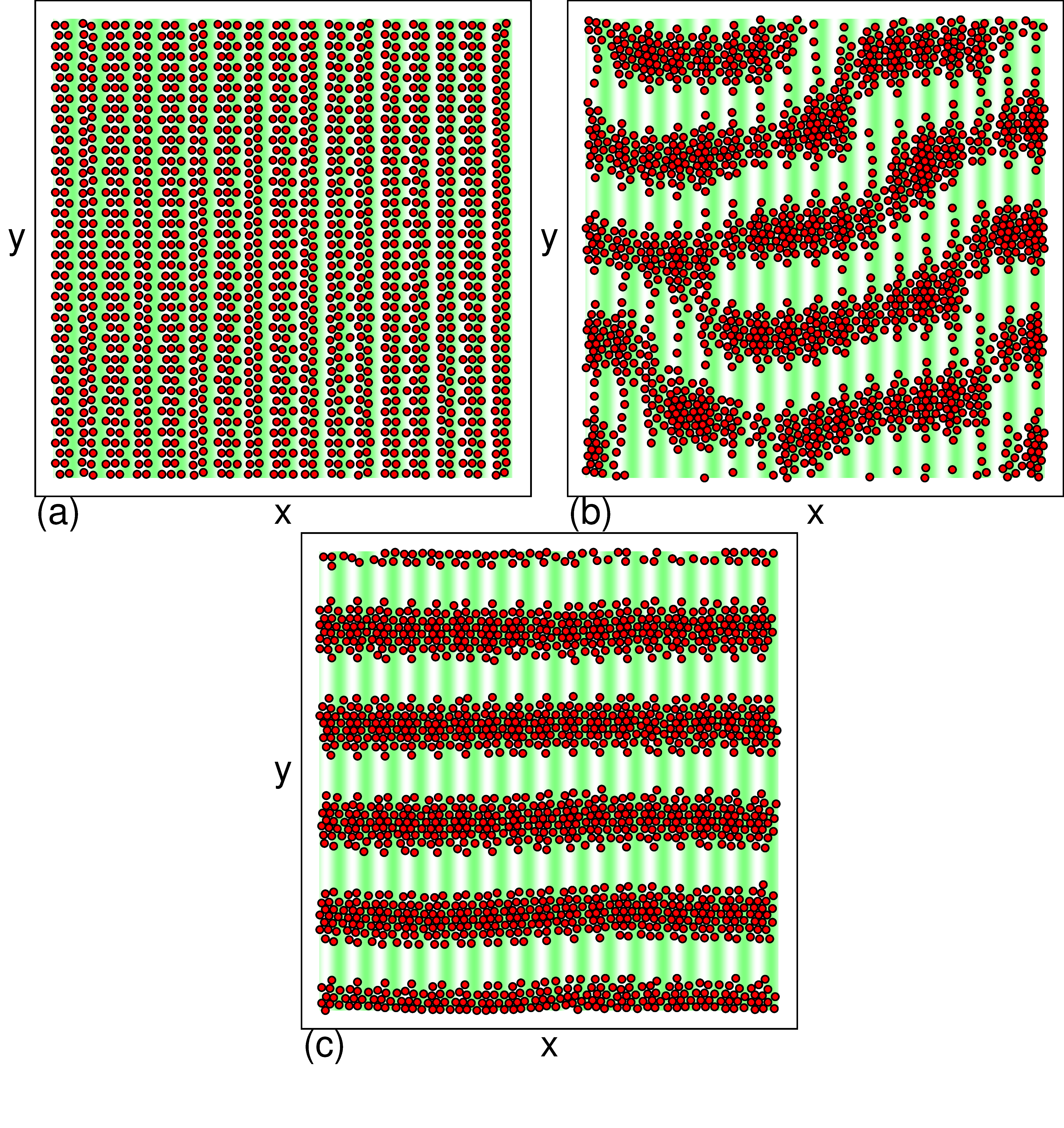}
\caption{Particle positions (red circles) and substrate maxima (green) and
minima (white) for the system from Fig.~\ref{fig:17} with
$a_p=2.067$, $B=2.25$, $F_p=1.25$, and $\rho=1.32$.
The circles representing the particles are drawn with a reduced radius
compared to the images obtained at lower $\rho$.
(a) $y$-aligned stripes at $\theta = 0.0^\circ$.
(b) A disordered stripe state at $\theta = 15^\circ$.
(c) $x$-aligned stripes at $\theta = 90^\circ$.
}
\label{fig:20}
\end{figure}

In Fig.~\ref{fig:20}(a), we show the particle configurations at
$\theta = 0.0^\circ$ for the $\rho = 1.32$ system from
Fig.~\ref{fig:17} where $y$-aligned stripes appear.
At $\theta=15^\circ$ in
Fig.~\ref{fig:20}(b), the aligned stripes break apart and the system
begins to assemble into disordered stripes that are partially aligned
in the $x$ direction.
For $\theta=90^\circ$ in
Fig.~\ref{fig:20}(c),
the stripes have become well aligned along the $x$ direction.
Even though a structural transition occurs in this system, there is
only weak hysteresis since the effectiveness of the pinning is
relatively weak.
For $\rho > 1.57$ (not shown), the system
fores a uniform crystal that undergoes almost no structural change
as the drive is rotated, and the hysteresis vanishes.

\section{Varied Substrate Spacing}

\begin{figure}
  \centering
  \includegraphics[width=\columnwidth]{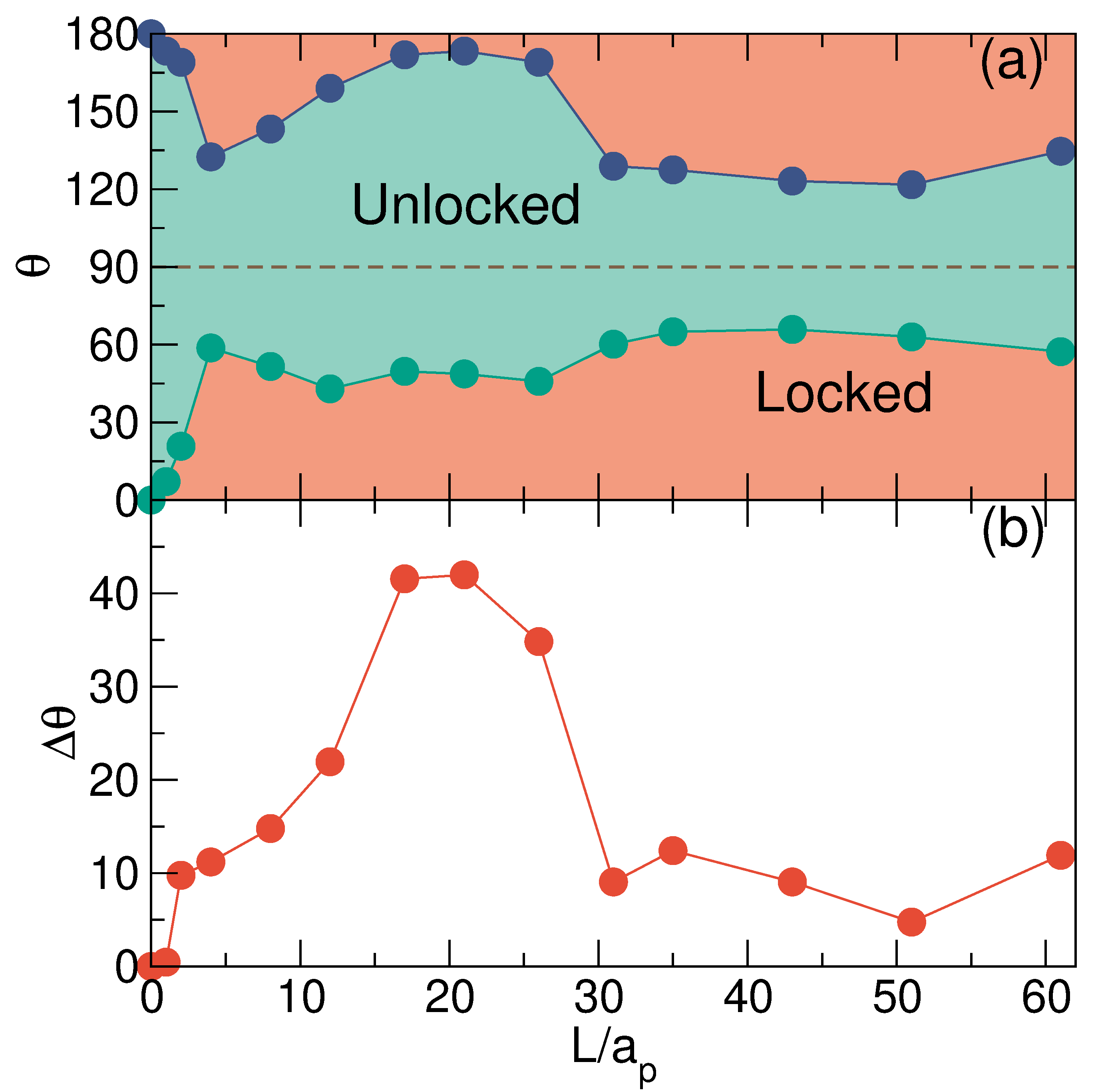}
\caption{(a) Boundaries $\theta_{\rm low}$ (green) and $\theta_{\rm hi}$ (blue)
  separating the regions where the motion is locked and not locked along the
  $\pm y$ direction plotted as a function of $\theta$ vs $L/a_p$
  for a system with $B=2.25$, $F_p=1.25$, and $\rho=0.454$.
 (b) The corresponding $\Delta \theta$ vs $L/a_p$.
}
\label{fig:21}
\end{figure}

In Fig.~\ref{fig:21}(a), we plot the boundaries
$\theta_{\rm low}$ and $\theta_{\rm hi}$ separating the regions of $y$-locked
motion from unlocked motion for a system with $F_p=1.25$ and $B=2.25$ as
a function of $\theta$ versus $L/a_p$, the number of substrate minima.
We note that up until this point, we have presented results
from systems where $a_p=2.067$ and $L/a_p=17$.
Figure~\ref{fig:21}(b) shows the corresponding $\Delta \theta$
versus $L/a_{p}$, where we find strong hysteresis for
$10 < L/a_p < 30$.
In the hysteretic regime,
the particles
form stripes or labyrinths that are aligned in
the $y$ direction, while
for larger values of $L/a_p$ corresponding to smaller values of $a_p$,
the system forms stripes that are aligned in
the $x$ direction. In the latter regime, 
fewer structural rearrangements occur as the drive is rotated, so the
hysteresis is reduced.
For $L/a_p < 10$ at $\theta = 0.0^\circ$, although the stripes
are aligned with the $y$ direction, each stripe is multiple particles
wide, which reduces the effectiveness of the
substrate pinning and
thus diminishes the amount of hysteresis that appears.

\begin{figure}
  \centering
  \includegraphics[width=\columnwidth]{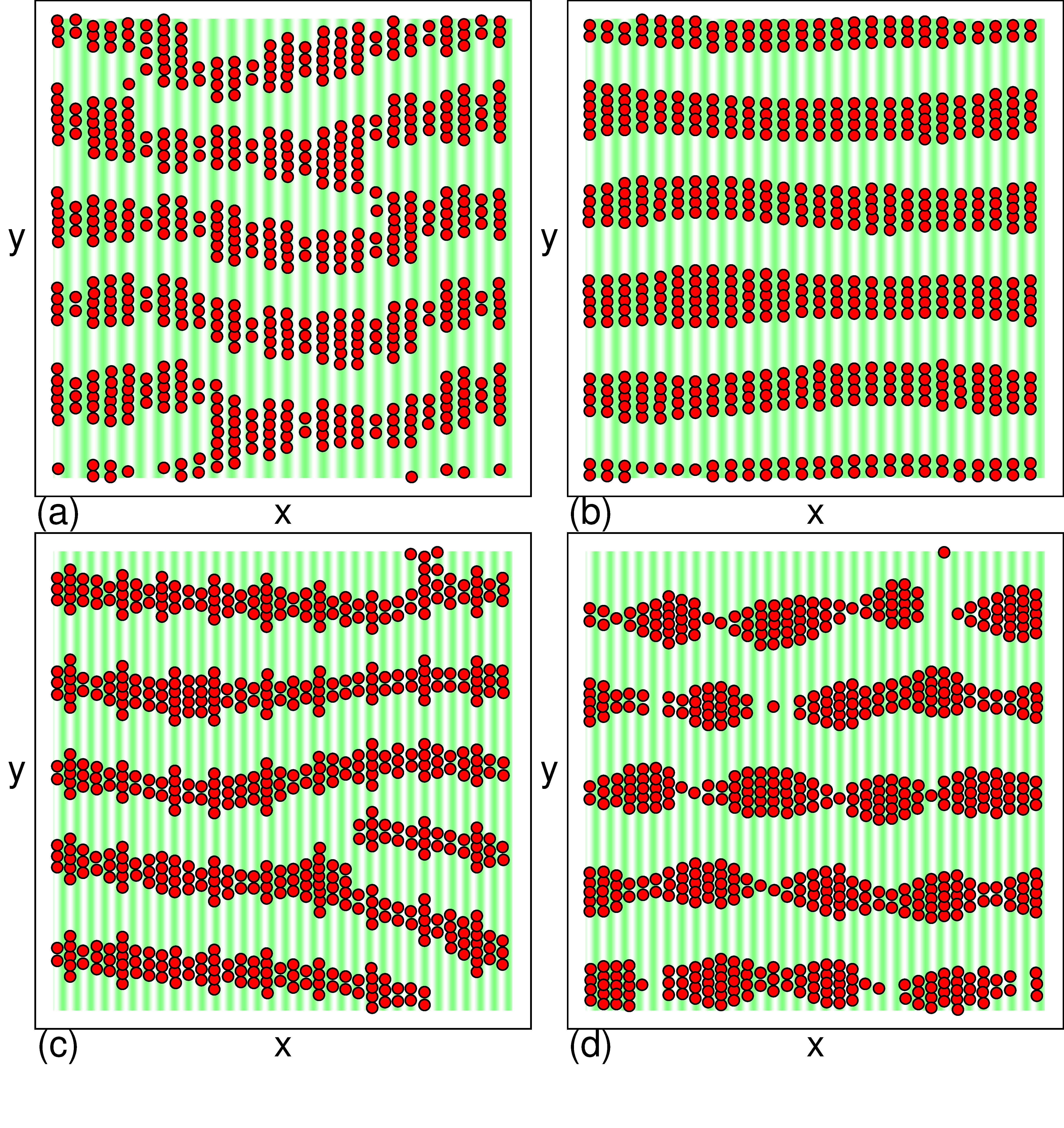}
\caption{Particle positions (red circles) and substrate maxima (green)
  and minima (white) for the system from Fig.~\ref{fig:21} with
  $B=2.25$, $F_p=1.25$, and $\rho=0.454$.
  (a) A labyrinthine pattern at $L/a_p=26$ and $\theta = 0.0^\circ$.
  (b) A stripe state at $L/a_p=26$ $\theta = 90^\circ$.
  (c) A stripe state at $L/a_p=35$ and $\theta = 0^\circ$.
  (d) A stripe state at $L/a_p=35$ and $\theta = 180^\circ$.
}
  \label{fig:22}
\end{figure}

In Fig.~\ref{fig:22}(a), we show a labyrinth-like phase that forms for
$L/a_p=26$ when $\theta=0.0^\circ$,
and in Fig.~\ref{fig:22}(b) we show a stripe state at
$\theta = 90^\circ$ for the same system.
For this substrate spacing, pronounced hysteresis arises since
the structure changes significantly
as the drive is rotated.
The ordered stripe phase illustrated in
Fig.~\ref{fig:22}(b) has $\theta_{\rm low}>180-\theta_{\rm hi}$, and
persists over a wider range of $\theta$ as the drive is
rotated into the $-y$ direction for $\theta>90^\circ$.
In Fig.~\ref{fig:22}(c), we show the stripe phase at
$\theta = 0.0^\circ$ for $L/a_p = 35$,
where the hysteresis is reduced.
In the same system at $\theta=180^\circ$, Fig.~\ref{fig:22}(d) indicates that
a stripe state is still present. Although some rearrangements of the
stripes occur as the drive is rotated, the general structure and orientation
of the stripes remains unchanged, reducing the amount of hysteresis that occurs.

\section{Summary} 

We have examined the symmetry locking and dynamic phases for
crystal, stripe, and bubble forming
systems interacting with a one-dimensional periodic substrate that has
its periodicity along the $x$ direction
under a drive that is rotated from the $+y$ direction to the $-y$
direction.
Here, the particle-particle interactions are composed of 
a combination of long-range repulsion and short-range attraction.
We measure the range of drive angles
over which the motion remains locked
along the $y$ direction,
and find that in regimes where stripes are present that are initially
aligned with the $y$ direction, the directional locking of the
motion under the rotating drive is the strongest.
The $y$-aligned stripes can dynamically reorder
into wider $x$-aligned stripes,
resulting in strongly asymmetric
velocity versus angle curves and a hysteretic response to the drive.
The asymmetry of the response persists even if multiple rotations
of the drive are performed,
and occurs because the $x$ alignment of the stripes
persists longer during the $-y$ phase of the drive rotation than during
the $+y$ phase of the drive rotation.
The hysteresis is generally reduced in the
bubble and crystal phases.
When the substrate is very strong,
the bubble phase can also show pronounced hysteresis when
the flow consists of a bubble-stripe mixture where the stripe portion
of the particles have their motion locked along the $y$ direction but
the bubble portion of the particles move in both the $x$ and $y$ directions.
In the crystal phases, the hysteresis is
always strongly reduced.
For higher particle densities,
the system can undergo structural transitions,
but the locking effects are strongly reduced when
the density becomes high enough to cause the stripes to grow
very wide.

\acknowledgments
We gratefully acknowledge the support of the U.S. Department of
Energy through the LANL/LDRD program for this work.
This work was supported by the US Department of Energy through
the Los Alamos National Laboratory.  Los Alamos National Laboratory is
operated by Triad National Security, LLC, for the National Nuclear Security
Administration of the U. S. Department of Energy (Contract No. 892333218NCA000001).

\bibliography{mybib}

\end{document}